\documentclass[preprint,authoryear,12pt]{elsarticle}

\usepackage{subcaption}
\usepackage{amsfonts}
\usepackage{amsmath}
\usepackage{amssymb}
\usepackage{graphicx}
\usepackage{geometry}
\usepackage{multirow}
\usepackage{booktabs}
\usepackage{adjustbox}
\usepackage{xcolor}

\journal{Transportation Research Part B}

\begin{document}

\begin{frontmatter}

\title{Origin-destination travel demand estimation: An approach that scales worldwide, and its application to five metropolitan highway networks}

\author[label1]{Chao Zhang\corref{cor1}}

\author[label1]{Neha Arora}

\author[label1]{Christopher Bian}

\author[label1]{Yechen Li}

\author[label1]{Willa Ng}

\author[label1]{Andrew Tomkins}

\author[label1]{Bin Yan}

\author[label1]{Janny Zhang}

\author[label1,label2]{Carolina Osorio}

\cortext[cor1]{Corresponding author. Email address: chaozhng@google.com. Postal address: 1600 Amphitheatre Pkwy, Mountain View, CA 94043.}

\address[label1]{Google Research, Mountain View, CA, USA}
\address[label2]{HEC Montr\'{e}al, Montr\'{e}al, QC, Canada}

\begin{abstract}
Estimating Origin-Destination (OD) travel demand is vital for effective urban planning and traffic management. Developing universally applicable OD estimation methodologies is significantly challenged by the pervasive scarcity of high-fidelity traffic data and the difficulty in obtaining city-specific prior OD estimates (or seed ODs), which are often prerequisite for traditional approaches. Our proposed method directly estimates OD travel demand by systematically leveraging aggregated, anonymized statistics from Google Maps Traffic Trends, obviating the need for conventional census or city-provided OD data. The OD demand is estimated by formulating a single-level, one-dimensional, continuous nonlinear optimization problem with nonlinear equality and bound constraints to replicate highway path travel times. The method achieves efficiency and scalability by employing a differentiable analytical macroscopic network model. This model by design is computationally lightweight, distinguished by its parsimonious parameterization that requires minimal calibration effort and its capacity for instantaneous evaluation. These attributes ensure the method's broad applicability and practical utility across diverse cities globally. Using segment sensor counts from Los Angeles and San Diego highway networks, we validate our proposed approach, demonstrating a two-thirds to three-quarters improvement in the fit to segment count data over a baseline. Beyond validation, we establish the method's scalability and robust performance in replicating path travel times across diverse highway networks, including Seattle, Orlando, Denver, Philadelphia, and Boston. In these expanded evaluations, our method not only aligns with simulation-based benchmarks but also achieves an average 13\% improvement in it's ability to fit travel time data compared to the baseline during afternoon peak hours.
\end{abstract}

\begin{keyword}
Origin-destination estimation; path travel times; travel behavior; optimization.
\end{keyword}

\end{frontmatter}

\section{Introduction}
Origin-Destination (OD) estimation is a cornerstone of transportation science, providing a fundamental building block upon which critical decisions in operations, policy, and planning rely. Estimates of travel demand (i.e., OD estimates) are indispensable for reliable traffic forecasting, efficient network management, strategic infrastructure investment, and the evaluation of mobility solutions. These estimates enable cities to improve quality of life, accessibility, equity, and sustainability goals set by transportation planners and engineers. For instance, understanding where trips start and end helps mobility planners improve transit offerings and run equity analyses to quantify how subpopulations of travelers benefit from, or are impacted by, specific infrastructure investments, changes in road-use pricing, road-use allocation strategies, as well as more strategic land-use policies.

Traditional OD estimation approaches most often depend on sparse segment count data and on periodic census or travel diary surveys, methodologies inherently limited in their spatial coverage, temporal granularity, and responsiveness to dynamic changes in travel behavior. The spatial sparsity of segment count data is due to the high costs of deploying and maintaining traditional loop-detector sensors. Similarly, the high cost of carrying out travel surveys, let alone census surveys, limits the ability to obtain recently collected data that represents well prevailing travel patterns. It is therefore difficult to collect such data across cities worldwide (i.e., at scale).  

This paper focuses on the design of OD estimation methods that can be readily applied at scale. We propose an approach that relies on travel time statistics. More specifically, we use ramp-to-ramp travel time data on highway networks. To the best of our knowledge, this work introduces the first scalable and analytical approach that uses path travel time data for OD estimation. Furthermore, this research is also the first work to develop a method that uses ramp-to-ramp travel times for all ramp pairs in a full metropolitan highway network. Most importantly, the proposed method does not rely on the existence of a seed OD (also known as a historical OD or a prior OD). The approach is formulated as a one-dimensional nonlinear constrained optimization problem, which embeds information from a fast-to-compute and differentiable analytical macroscopic network model. This model requires limited data for calibration purposes. The need for limited data needed to calibrate the model makes the approach scalable worldwide. It also makes the approach suitable for large-scale network analysis. This is illustrated by applying the approach to the full metropolitan highway networks of seven major US cities (two for validation, five for case studies).

The context of this paper is on the estimation of ODs (also known as trip matrices) using ramp-to-ramp route travel times for metropolitan-scale highway networks. A path consists of a ramp-to-ramp route, always beginning with an on-ramp and ending with an off-ramp. We assume that such path travel times are observed for a subset of vehicles from the entire population of vehicles.

Since the pioneering OD estimation works, e.g., \cite{Cascetta13}, the most common type of ground truth (GT) (i.e., field) data has been the use of spatially sparse segment counts. This remains true, as noted by \cite{Castillo15}. However, there is an increasing focus on the use of other types of data and various sensors. An interesting review of past work that uses trajectory data to infer traffic states (not limited to OD estimation) is given in Table 1 of \cite{Yu20}. We focus here on work that relies on path travel time data obtained from trajectory datasets. Path travel times can be measured through Automatic Vehicle Identification (AVI) sensors, such as license plate recognition sensors. These sensors record travel times for a subset of the vehicle population. The assumption that only a subset of vehicles is observed can be due to sensors only covering a subset of lanes of the segments equipped with the sensors, e.g., \cite{Asakura00}, or due to the sensors only observing vehicles equipped with AVI tags, e.g., \cite{Zhou06}.

The reliability of AVI measurements is known to vary with levels of congestion, weather, and daylight \citep{Asakura00, Castillo13b}. These challenges may partially explain why the OD estimation literature with access to AVI data relies most often on the use of point-to-point count data \citep{Asakura00, Castillo13a} or point-to-point split ratio data \citep{Zhou06} rather than point-to-point travel time data. Instead of using the travel time data as calibration data (i.e., using it as the field data that the estimation process aims to  replicate), it is most often used to estimate other model parameters. For instance, \cite{Dixon02} use it to estimate link choice probabilities for real-time OD estimation.
Another reason why path travel time data may not have been used as calibration data even when available, is likely the fact that it introduces nonlinearities in the problem formulation. For instance, the work of \cite{Xing24} has access to taxi probe path travel time data, which is used to estimate segment travel time data, followed by segment count data; it is the count data that is used as calibration data. This allows to assume a linear, or in some cases piece-wise linear, relationship between the ODs and the calibration data. If travel time data were used, instead of counts, this relationship would become nonlinear. Similarly \cite{Aslam12} study the properties of the probe OD obtained from a probe taxi fleet, yet they do not directly use the path travel time data to improve or expand the probe OD. \cite{Mohanty20} tackle the dynamic OD estimation problem using, among others, path travel time estimates available from navigation application interfaces to infer route choice probabilities. Again travel time data is not directly used for calibration purposes. The work of \cite{Asakura00} has access to travel time data from AVI sensors, but limits its use to validation purposes. Instead, they use partial counts (i.e., counts of the probe vehicle population) data for calibration.

The work of \cite{Asakura00} is similar to ours in that it proposes an analytical optimization approach for static OD estimation on highways, the method is formulated as a constrained least squares optimization problem and considers a single route per OD. Like our work, they have access to travel time data from AVI sensors, however they do not directly calibrate travel times and instead use the travel time data for validation purposes. They use partial counts (i.e., counts of the probe vehicle population) data for calibration. Other ways in which the work differs from our work is in that the formulation is  applicable only to a single highway corridor and is illustrated with a Kobe (Japan) highway corridor case study, while our approach can be applied to a large-scale highway network with an arbitrary network topology, it is not limited to a single corridor layout. As is common in the OD literature, \cite{Asakura00} assume knowledge of a prior OD; however, we do not make this assumption.  

Given the need for a scalable approach, we do not assume that third party data, such as prior population ODs are available. This is due to the difficulty of obtaining such data \citep{Mohanty20}, let alone obtaining it at scale, and to the challenges of quantifying the reliability of such data. For instance, it is often outdated, in particular in light of the significant changes in travel patterns observed post-COVID. Unlike the proposed approach, most OD estimation work assumes the existence of a prior population OD, typically obtained from travel or census surveys. For instance, the work of \cite{Toole15} uses socioeconomic census data, for the entire population, to upsample a probe OD obtained from call detail record data. Our proposed approach assumes the availability of OD level aggregated statistics (e.g., counts, travel times), we then upsample the counts using travel time data. Hence, we do not require any prior population-level (e.g., census) data.

To more accurately describe the temporal variations of congestion patterns, research has focused on tackling the OD estimation problem through the use of time-dependent network loading models that are implemented as simulators, rather than analytical models. Examples include the use of mesoscopic traffic models \citep{Zhou06, Cipriani14, Zhang24}. Similarly, macroscopic traffic models defined as partial differential equations have been used, they are evaluated using tailored numerical computation schemes, often in the form of finite-difference approximations or event-based simulation processes. For instance, Newell’s simplified kinematic wave model \citep{Newell93} is used in \cite{lu13}, and the  Lighthill-Whitham-Richards (LWR) model \citep{Lighthill55,Richards56} is used in \cite{Qian11}. The context of this paper is the need to develop a technique that scales globally. This need for scalability implies that the proposed OD estimation method is to be applied for an arbitrary number of cities, and for every hour of an arbitrary number of days. Hence, a method with low compute times is needed such that it can be readily applied. For this reason, we opt to use an analytical, and differentiable, network loading model that is formulated as a system of nonlinear equations. It can therefore be evaluated instantaneously. This simplicity comes at the cost of reduced realism in describing congestion patterns. In particular, the proposed model assumes a stationary regime and does not describe the temporal variations of congestion patterns within the time interval of interest (e.g., the hour of the day for which the OD is estimated). 

Formulations with analytical network loading models from the literature can be extended to account for our current setting. For instance, the formulation of \cite{Castillo13a} consists of a least squares objective function subject to linear flow conservation constraints and non-negativity path flow bound constraints. It assumes the existence of a prior OD for the population of travelers. This model can be extended to account for the use of path travel time data and for the existence of a prior OD for a subset of travelers. Such an extension would convert the quadratic objective function into a polynomial of power four, and would integrate additional nonlinear equality constraints. Similarities between our work and that of \cite{Zhan13} are also of interest. They study the inverse problem of ours, but have similarities in the formulation. They assume the OD is known and probe subpath travel time data is available. They propose a least squares formulation with nonlinear equality constraints and non-negativity constraints to infer full path travel times.

There is abundant OD estimation literature that computes equilibrium conditions. This leads to  formulations as bi-level optimization problems or variational inequalities. Our focus is on highway networks, the set of origins (resp. destinations) consists of all highway on-ramps (resp. off-ramps). For each on-ramp off-ramp pair, we assume there is a single route. This is not a strong assumption in a highway network setting. This assumption allows us to avoid the modeling of equilibrium conditions. However, this assumption can be readily relaxed in various ways. If reliable route choice probabilities can be estimated from aggregated statistics, then the assumption can be relaxed without the need to include equilibrium constraints. Alternatively, the formulation can be extended to include nonlinear route choice probability equality constraints, as Eq.~(7) of \cite{osorio19b}, that model stochastic user equilibrium conditions and remain numerically tractable to evaluate. 

The remainder of the manuscript is structured as follows. Section~\ref{sec:method} formulates the proposed OD estimation method. Section~\ref{sec:valid} validates the proposed approach using field segment count data for the highway networks of San Diego and Los Angeles. Section~\ref{sec:casestudies} illustrates the scalability of the approach by applying it to the highway networks of Seattle, Orlando, Denver, Philadelphia, and Boston. Section~\ref{sec:cl} discusses the findings and presents concluding remarks.

\section{Methodology}
\label{sec:method}
We aim to estimate an OD matrix for vehicular traffic for a metropolitan highway network for a specific time interval of interest (e.g., a given hour). We utilize two sets of aggregated, anonymized OD statistics from Google Maps Traffic Trends: (1) average travel times (also referred to as ground truth or GT) and (2) subsample ramp-to-ramp traffic counts. The goal is to estimate an OD demand based on these ramp-to-ramp counts that, when loaded on the network, is expected to replicate the GT travel time patterns. 

The OD estimation problem is known to be underdetermined, also known as ill-specified or ill-posed. This means that there are multiple ODs, and often an infinite number, that when loaded onto the network would equally fit the traffic data. The level, or amount, of underdetermination is greater when spatially sparse data, such as the most commonly used segment count data, is used. The level is significantly lower when using data with broader spatial coverage, such as travel time statistics. However, the problem remains underdetermined.

Given this underdetermination, it is important to impose a certain structure on the desired ODs so as to yield an OD that would be plausible for the considered city and time interval. Traditional methods impose structure by assuming the availability of a city-provided seed OD matrix of the population. Since we aim to develop a method that can be applied worldwide without the need for collecting additional data, we do not assume availability of a city-provided seed OD matrix. Instead, we use an OD matrix estimated from ramp-to-ramp traffic count statistics. We call this matrix the subsample OD. However, this matrix represents only a sample of the vehicular population and therefore needs to be upscaled. For a given time interval, we uniformly (across space) upscale the subsample OD matrix. The scaling factor is determined by solving an optimization problem that embeds an analytical and differentiable network model, approximating path travel times from a given upscaled OD matrix. The approach preserves subsample demand structure while being numerically fast to solve and not requiring additional city-provided OD or sensor segment count data.

To formulate the method, we use the following notation.\\
\begin{tabular}{l}
\textbf{Endogenous model variables:}\\
$x$: scalar demand scaling factor;\\
$k_i$:  per-lane density of segment $i$;\\
$\lambda_i$:  travel demand of segment $i$;\\
$v_i$:  (space-mean) speed of segment $i$;\\
$t_p$:  travel time of path $p$.\\

\textbf{Exogenous parameters (i.e., model inputs):}\\
$a_{ij}$: element $(i,j)$ of the assignment matrix $A$ (i.e., network loading mapping);\\
$\alpha_1, \alpha_2$: power scalar coefficients of the segments' fundamental diagram;\\
$d_j$: demand of OD pair $j$ obtained from the subsample OD;\\
$\kappa$: demand to density scaling factor;\\
$k^{\text{jam}}$: segment jam density;\\
$l_i$: length of segment $i$;\\
$n_i$: number of lanes of segment $i$;\\
$t_p^{\text{GT}}$: ground truth (i.e., field data) travel time of path $p$;\\
$v^{\text{max}}_i$: maximum segment speed (e.g., free-flow speed);\\
$v^{\text{min}}_i$: minimum segment speed;\\
$w_p$:  weight parameter for the path travel time error term of path $p$;\\
$x_L, x_U$: scalar lower and upper bound values;\\

$\mathcal{I}$: set of network segment indices;\\
$\mathcal{I}(p)$: set of indices of segments of path $p$;\\
$\mathcal{J}$: set of OD pair indices;\\
$\mathcal{P}$: set of indices of paths with ground truth (GT) data available.\\
\end{tabular} \\

\noindent The method is formulated as follows.
\begin{align}
\min_{x} \quad & 
f(x) = \frac{1}{|\mathcal{P}|}\sum_{p\in\mathcal{P}}{w_p(t_p^{\text{GT}}-t_p)^2} \label{eq:objfn}\\
\textrm{s.t.} \quad &  \lambda_i = x \left( \sum_{j \in \mathcal{J}} a_{ij} d_j \right) \quad  \forall i \in \mathcal{I} \label{eq:lambda}\\
\quad & k_i = \frac{\kappa k^{\text{jam}}}{n_i} \lambda_i \quad \forall i \in \mathcal{I} \label{eq:k}\\
\quad & v_i = v^{\text{min}}_i + (v^{\text{max}}_i - v^{\text{min}}_i)\left(1 - \left(\frac{k_i}{k^{\text{jam}}}\right)^{\alpha_1}\right)^{\alpha_2}\forall i \in \mathcal{I} \label{eq:v}\\
\quad &  t_p = \sum_{i\in \mathcal{I}(p)} \frac{l_i}{v_i} \quad \forall p \in \mathcal{P} \label{eq:t}\\
\quad & x_L \le x \le x_U \label{eq:bounds}.
\end{align}
The objective function~\eqref{eq:objfn} considers a scalar decision variable $x$ and uses a quadratic function to measure the distance between the model-based path travel times $t_p$ and their data-based ground truth equivalents $t_p^{\text{GT}}$. This is a commonly used distance metric in the OD estimation literature. The error term of each path is weighted by scalar weights ($w_p$). These can be used to, for example, emphasize the error of longer paths, paths that are more frequently traveled, or paths with higher levels of congestion. 

Eq.~\eqref{eq:lambda} is a linear network loading equation. Entry $a_{ij}$ of an assignment matrix $A$ represents the probability that demand of OD pair $j$ travels on segment $i$. The system of linear and nonlinear equality constraints Eqs.~\eqref{eq:k}-\eqref{eq:t} represent a network model which builds upon the model of \cite{osorio19b} (Eqs.~(6)-(11) in that paper). The main distinctions are that the proposed approach does not account for route choice because in this case all ODs have a single route, and the fundamental diagram (FD) has been updated to allow for a non-zero minimum speed ($v^{\text{min}}_i$). Eq.~\eqref{eq:k} linearly maps segment demand to lane density. Here, $k_i$ is a per-lane density for segment $i$. It assumes that all lanes of a segment have common density. Eq.~\eqref{eq:v} is the nonlinear FD of the segment. This is a commonly used, and broadly validated, FD formulation for highways. However, if the same travel time data for full metropolitan networks (including urban and arterial roads) were to be used, the FDs would need to be updated. For instance, recent progress on the shape of FDs for urban and signalized segments would be appropriate \citep{Li22,Zhang22fd}. We use the same FD functional form for all segments in the network. Given our focus on using highway data, we consider a unique jam density value ($k^{\text{jam}}$ of Eq.~\eqref{eq:k}) for all segments. Similarly, we assume that all segments share the same power coefficients for their FD ($\alpha_1$ and $\alpha_2$ of Eq.~\eqref{eq:v}). These assumptions can be readily relaxed. Their relaxation would merely increase the number of input parameters to estimate. Eq.~\eqref{eq:t} defines the path travel time as the sum of the segment travel times for all segments on the path. Eq.~\eqref{eq:bounds} defines upper and lower bounds for the demand scaling factor (i.e., the decision variable). 

This formulation consists of a one-dimensional optimization problem with a quadratic objective function, and nonlinear equality constraints. Derivatives of all equality constraints are explicitly computed, and the formulation is implemented as a problem with nonlinear (and non-quadratic) objective function constrained only by bounds. 

We consider a single upscaling factor $x$ (i.e., $x$ is a scalar rather than a vector). This means that all OD entries are upscaled by the same multiplier. This is equivalent to assuming that the derived subsample OD is a uniform sample from the entire population of travelers. This may be a strong assumption depending on how the subsample OD is collected. This assumption can be relaxed, as desired, by converting $x$ from a scalar into a vector such as to upscale subsets of OD entries equally, or ultimately having a single scalar for each OD entry. The latter approach is the most flexible but it does not preserve any structure from the subsample OD, hence another way to regularize the problem and to enforce structure would need to be added. 

In the proposed approach, we discretize time (e.g., on an hourly basis) and apply the method, separately, for each discrete time interval. For a given time interval, the decision variable of our problem $x$ can be interpreted as the market penetration rate. The difficulty of estimating market penetration rates or identification rates of AVI or other probe sensors is well discussed in the literature \citep{Zhou06}. The proposed work allows these rates to vary across time intervals and, similar to past work \citep{Dixon02}, assumes that the rates do not vary over space. This assumption leads to an estimated population OD that preserves the structure in the subsample OD. This assumption can be readily relaxed to consider a specific spatial variation of the rate, if such information is available. 

Since the focus is on highways, the OD matrix defines origins (resp. destinations) as highway entry (resp. exit) points. In this case the notion of path travel time coincides with that of OD travel times. In other words, we do not consider travel times of subpaths. Hence, the sets $\mathcal{J}$ and $\mathcal{P}$ are identical. In the case studies of this paper, we weigh all error terms equally in the objective function, i.e., in Eq.~\eqref{eq:objfn} $w_p=1 \quad \forall i \in \mathcal{P}$.

\section{Validation}
\label{sec:valid}
This section validates the proposed approach. The goal is to quantify the ability of the estimated ODs to reproduce congestion patterns observed in the network. Recall that we estimate the ODs using travel time data. To carry out a robust validation, we use a different type of data: segment count data. The model is not designed to replicate segment count data, nor does it use any segment count data in the estimation process. This ensures that the validation provides a robust assessment of the estimated OD demand's capability to replicate traffic patterns. 

We consider two large-scale highway networks: Los Angeles and San Diego. We use segment traffic count data obtained from roadway sensors deployed across the networks. The data is obtained from the California Department of Transportation portal and is a commonly used data source known as the Performance Measurement System (PeMS) dataset \citep{Chen01}. The proposed model approximates the traffic counts of a given segment $i$ as the product of the segment's number of lanes, lane-density, and speed: $n \times k \times v$, as defined in Eqs.~(3)-(4). We compare the performance of the proposed approach to that of a baseline that uses OD estimates derived directly from the unscaled subsample OD count statistics. This corresponds to an OD with no upscaling: $x=1$.

\begin{figure}[htbp]
    \centering
    \begin{subfigure}[b]{0.45\textwidth}
        \centering
        \includegraphics[width=\linewidth]{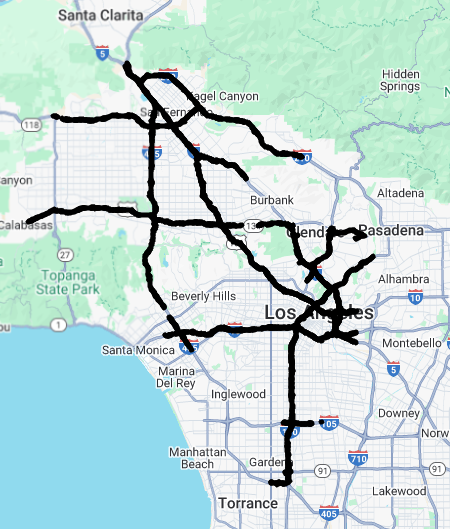}
        \caption{Los Angeles Highway Network.}
        \label{subfig:la_network}
    \end{subfigure}
    \hfill 
    \begin{subfigure}[b]{0.46\textwidth}
        \centering
        \includegraphics[width=\linewidth]{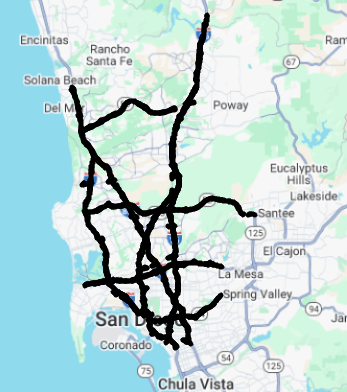}
        \caption{San Diego Highway Network.}
        \label{subfig:sd_network}
    \end{subfigure}
    \caption{Validation highway networks. The images are screenshots from Google Maps.}
    \label{fig:validation_networks}
\end{figure}

The highway networks for validation are visually presented in Figure \ref{fig:validation_networks}, with Table \ref{tab:validation_network_spec} providing their quantitative characteristics, such as the number of network segments and the maximum OD dimensionality observed during evaluation, across all hours, for a city. We use data from 11 sensors for Los Angeles, and from 17 sensors for San Diego. For each sensor, we use hourly segment count data for a weekday from 5 AM to 8 PM. This includes both off-peak and peak traffic conditions. It is worth noting that the Los Angeles network is known for its notoriously high levels of congestion, so this presents a particularly challenging validation scenario. 

\begin{table}[htbp]
    \centering
    \caption{Validation network specifications.}
    \begin{tabular}{ccc} 
        \toprule
        \textbf{Network} & \textbf{Number of segments} & \begin{tabular}{@{}c@{}}\textbf{Max hourly}\\\textbf{OD dimensionality}\end{tabular} \\
        \midrule
        Los Angeles & 16,236 & 1,838 \\
        \midrule
        San Diego & 10,370 & 1,125 \\
        \bottomrule
    \end{tabular}
    \label{tab:validation_network_spec}
\end{table}

\begin{figure}[htbp]
    \centering
    \begin{subfigure}[b]{0.48\textwidth}
        \centering
        \includegraphics[width=\linewidth]{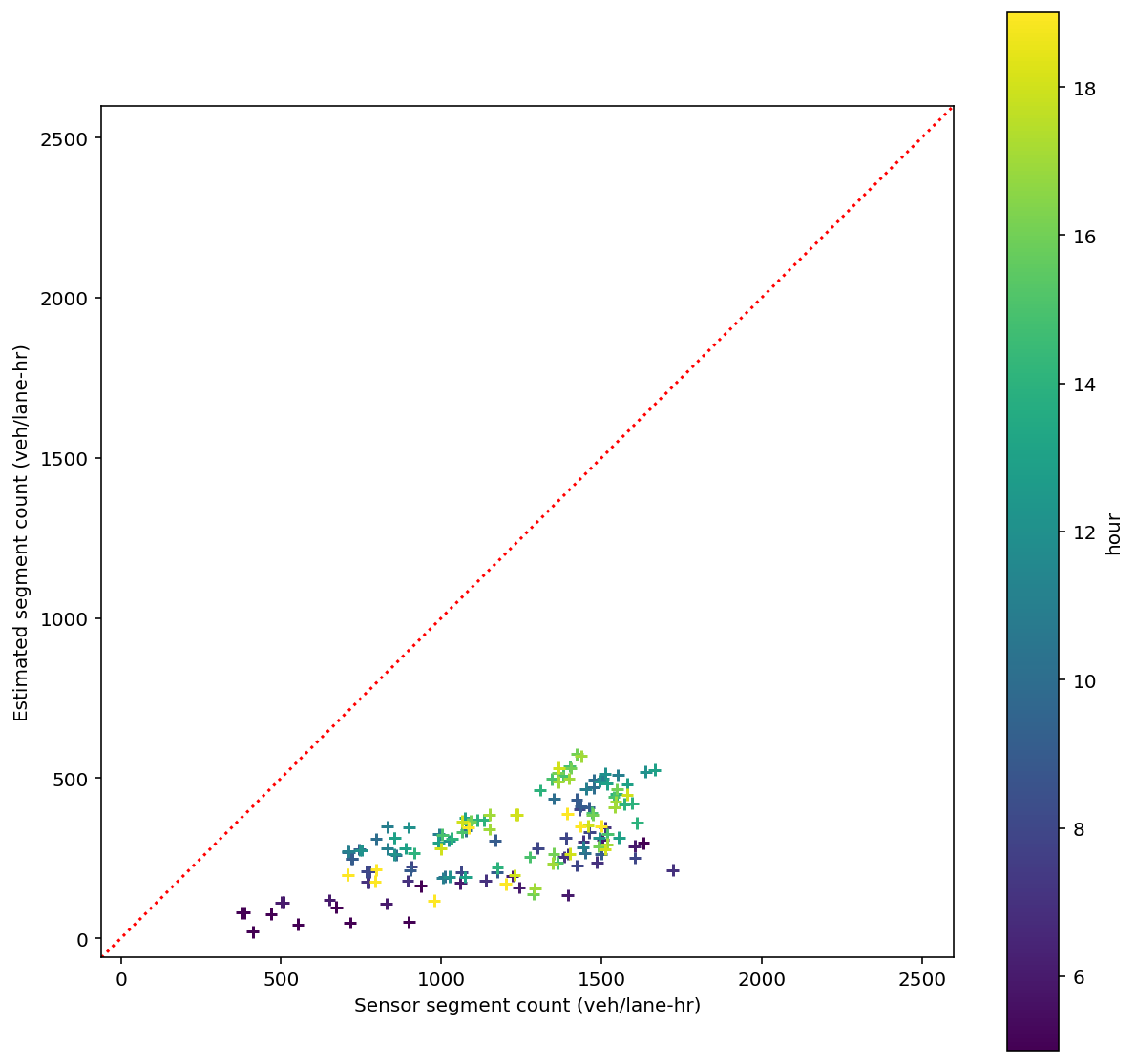}
        \caption{Los Angeles: Baseline.}
        \label{subfig:la_count_scatter_baseline}
    \end{subfigure}
    \hfill 
    \begin{subfigure}[b]{0.48\textwidth}
        \centering
        \includegraphics[width=\linewidth]{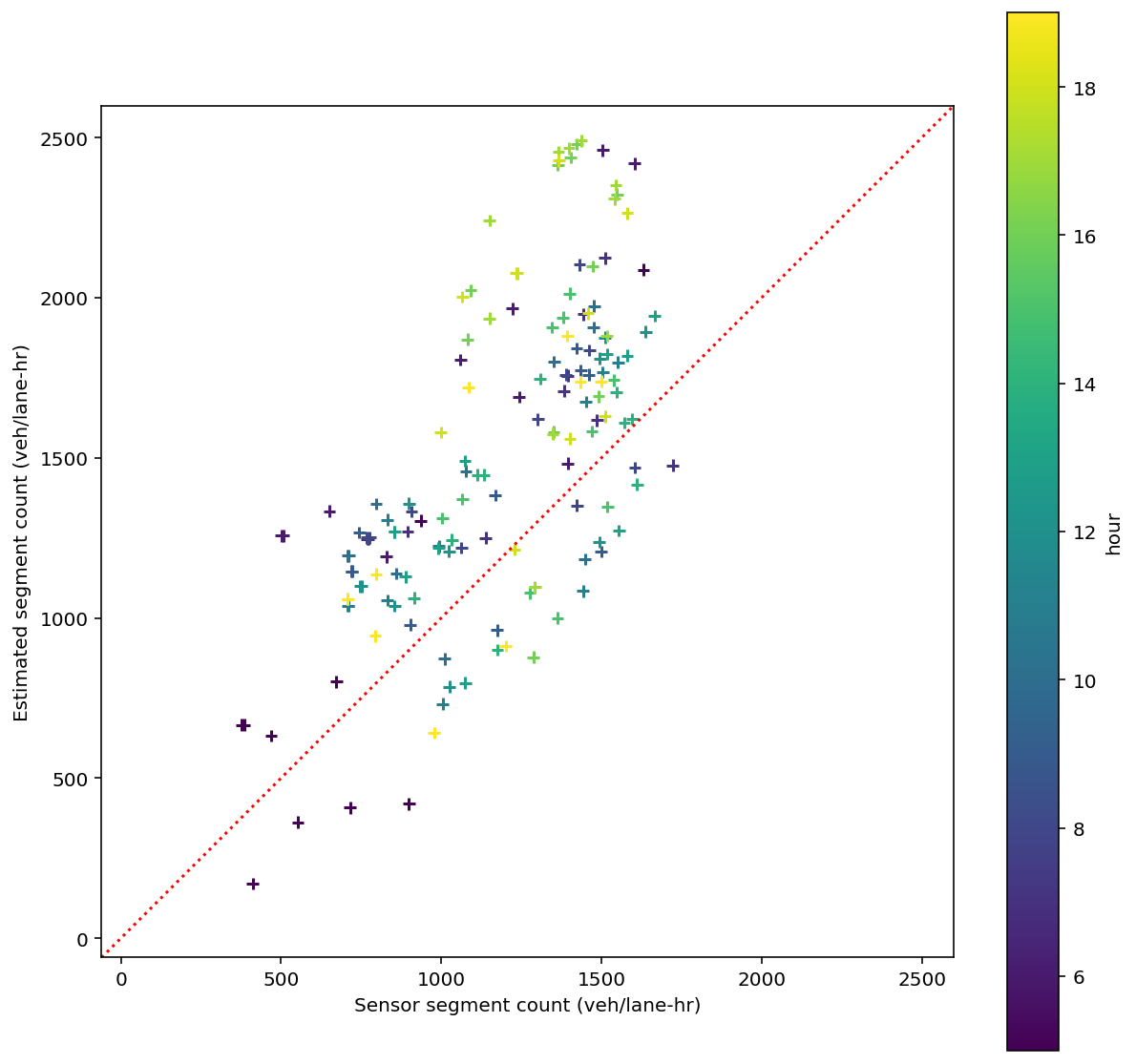}
        \caption{Los Angeles: Proposed Method.}
        \label{subfig:la_count_scatter_proposed}
    \end{subfigure}
    \caption{Comparison of segment counts from the baseline and proposed models against field data for the Los Angeles network.}
    \label{fig:la_validation_count_scatter}
\end{figure}

\begin{figure}[htbp]
    \centering
    \begin{subfigure}[b]{0.48\textwidth}
        \centering
        \includegraphics[width=\linewidth]{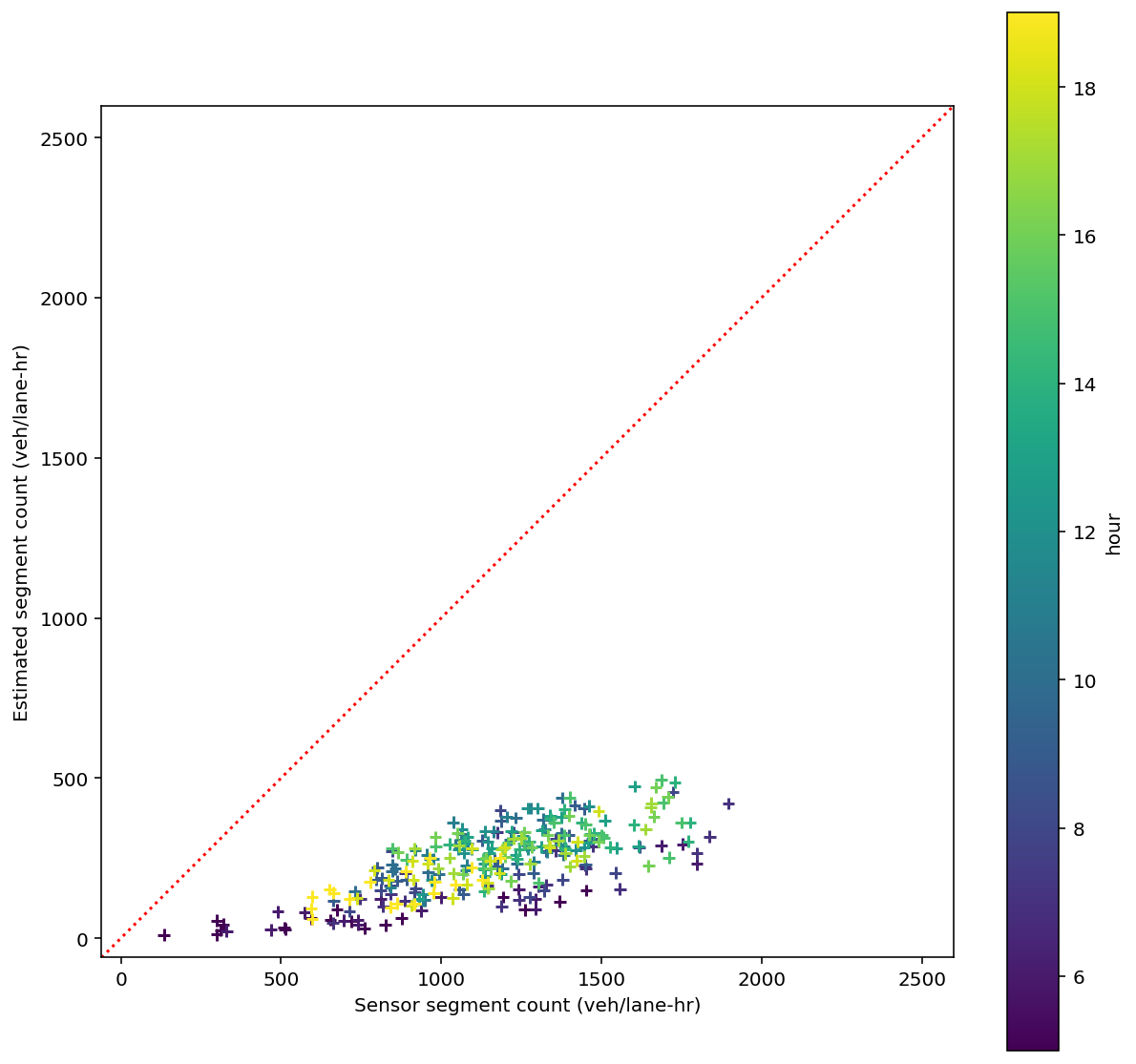}
        \caption{San Diego: Baseline.}
        \label{subfig:sd_count_scatter_baseline}
    \end{subfigure}
    \hfill 
    \begin{subfigure}[b]{0.48\textwidth}
        \centering
        \includegraphics[width=\linewidth]{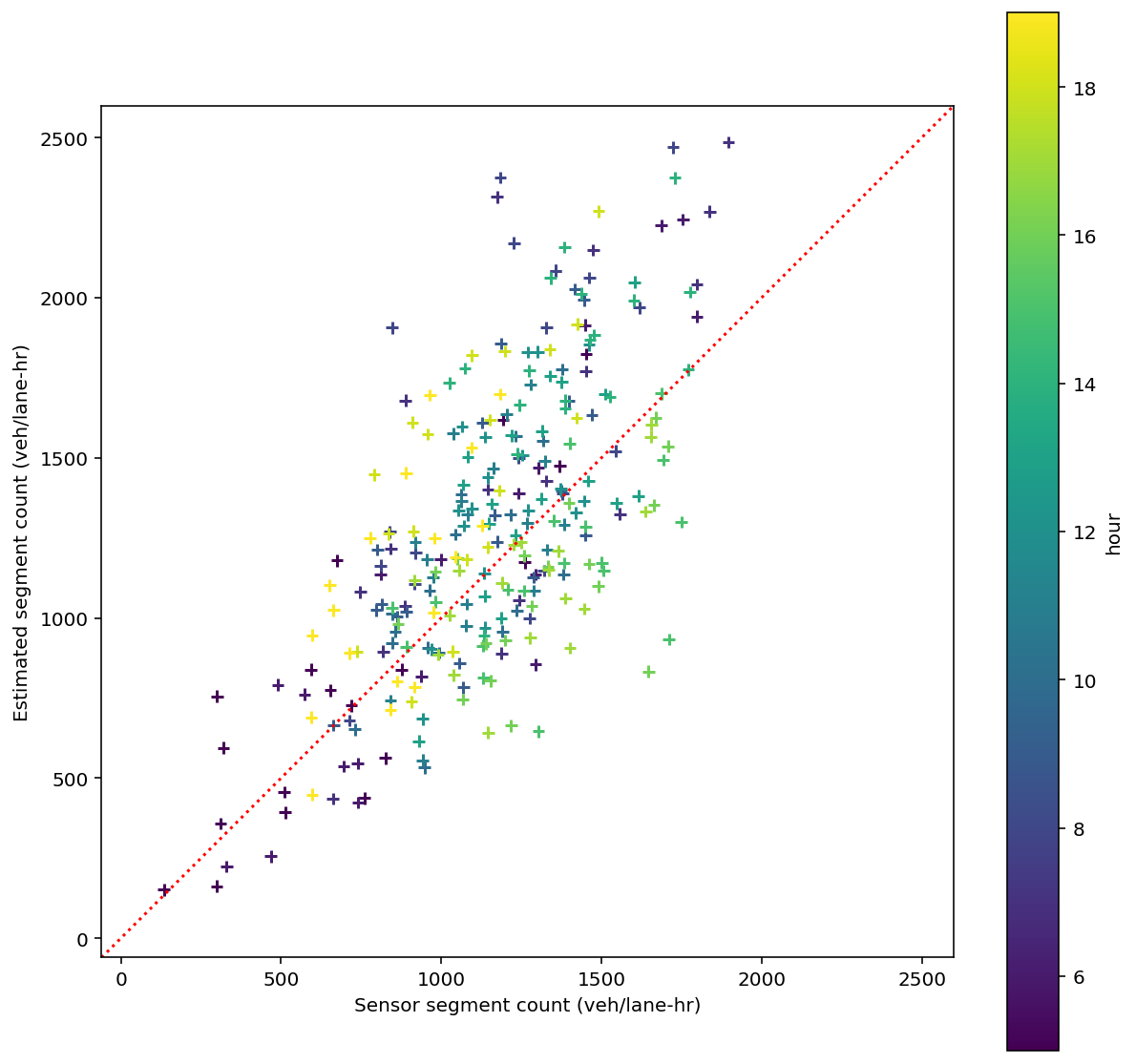}
        \caption{San Diego: Proposed Method.}
        \label{subfig:sd_count_scatter_proposed}
    \end{subfigure}
    \caption{Comparison of segment counts from the baseline and proposed models against field data for the San Diego network.}
    \label{fig:sd_validation_count_scatter}
\end{figure}

Figures \ref{fig:la_validation_count_scatter} and \ref{fig:sd_validation_count_scatter} visually assess the performance of our proposed model against an unscaled baseline by presenting scatter plots comparing ground truth (\textit{x}-axis) and estimated (\textit{y}-axis) hourly segment counts for each city. For each plot, two subplots are presented side-by-side, offering a direct comparison between our proposed model and the baseline. Both subplots maintain identical value ranges for their \textit{x} and \textit{y}-axes to ensure a direct and unambiguous visual comparison between the two. Each data point corresponds to an hourly segment count, which has been normalized by the number of lanes. The observations are distinguished by the hour of the day using a color-coding scheme, as illustrated in the accompanying color bar.

A key indicator of accuracy is the proximity of data points to the dotted identity line, which represents perfect agreement with ground truth. The plots clearly show that data points from our proposed method consistently cluster tightly around the identity line, signifying high estimation accuracy. This holds true and remains consistent across both cities and for all hours, from 5 AM to 8 PM, associated with diverse congestion levels. In stark contrast, the unscaled baseline exhibits a distinct and systematic downward shift, with points consistently falling significantly below the identity line. This substantial deviation indicates a pervasive underestimation of counts by the baseline.

Additionally, the validation quality is assessed in terms of the normalized root mean square error (\textit{nRMSE}) of measurements (in this case, counts) across all sensors over a given time range (e.g., across all time periods, or per time period), where \textit{nRMSE} is defined by Eq. \eqref{eq:nRMSE}:

\begin{align}
nRMSE = \frac{|\mathcal{S}|}{\sum_{s \in \mathcal{S}} y^{GT}_s} \sqrt{\frac{1}{|\mathcal{S}|}\sum_{s \in \mathcal{S}} \left(\hat{y}_s - y^{GT}_s\right)^2} \times 100,
\label{eq:nRMSE}
\end{align}
where $\mathcal{S}$ represents the collection of entities for which measurements are taken, such as network segments and/or time intervals. Specifically, $y^{GT}_s$ refers to the actual (ground truth) measurement for an entity $s \in \mathcal{S}$, while $\hat{y}_s$ indicates the estimated measurement for that same entity. In this section, \textit{nRMSE} is calculated for segment counts, and in Section~\ref{sec:casestudies}, it is calculated for path travel times. Throughout all the tables, \textit{nRMSE} values are presented as percentages, calculated using Eq. \eqref{eq:nRMSE} provided above (which includes a multiplier of $100$). Therefore, the $\%$ symbol is usually omitted for simplicity.

To further assess the comparative efficacy of different methodologies, we complement the use of \textit{nRMSE} with its percentage improvement (also referred to as `\% Improvement'). This metric provides a direct quantification of the relative enhancement in quality, thereby highlighting the impact of methodological refinements. The percentage improvement in \textit{nRMSE} is formally defined as:
\begin{equation}
\% \text{Improvement} = \frac{nRMSE_{\text{baseline}} - nRMSE_{\text{model}}}{nRMSE_{\text{baseline}}} \times 100,
\label{eq:percentage_improvement_nRMSE}
\end{equation}
where $nRMSE_{\text{baseline}}$ represents the \textit{nRMSE} of the baseline, while $nRMSE_{\text{model}}$ denotes the \textit{nRMSE} of a given method under evaluation, e.g., our proposed model, a benchmark. For conciseness, this metric is also presented as a percentage, with the $\%$ symbol omitted.

Tables \ref{tab:count_nrmse} and \ref{tab:hourly_count_nrmse} present a quantitative comparison of the segment count \textit{nRMSE} for our proposed methodology against the baseline. To provide an overview of the results, we first present a table summarizing the overall performance metrics (Table~\ref{tab:count_nrmse}). These metrics are aggregated for each city, encompassing data from all sensors and across all hours. Following this aggregated summary, a more detailed analysis is presented, with hour-specific metrics presented to illuminate temporal patterns and variations (Table~\ref{tab:hourly_count_nrmse}).

Table \ref{tab:count_nrmse} summarizes the overall \textit{nRMSE} statistics, aggregated across all hours for each city. Our proposed method achieves segment count \textit{nRMSE} values of 39\% for Los Angeles and 29\% for San Diego. These figures represent substantial improvements over the baseline, corresponding to reductions in count \textit{nRMSE} of 64\% for Los Angeles and 74\% for San Diego, respectively.

Table \ref{tab:hourly_count_nrmse} provides a more granular, hour-by-hour analysis of count \textit{nRMSE} across all sensors. In both validation networks, our proposed method consistently yields lower count \textit{nRMSE} values for every hour compared to the baseline, with the specific percentage improvement detailed in the table's rightmost column. While the proposed method demonstrates enhanced performance overall, some hourly variations are observed. For instance, in Los Angeles, improvements are less pronounced (i.e., under 50\% in count \textit{nRMSE} \% Improvement) during the morning (6-7 AM) and afternoon (4-7 PM) commute periods. San Diego shows a similar trend, with slightly diminished performance gains during its morning peak congestion hours (7-8 AM). 

\begin{table}[htbp]
    \centering
    \caption{Overall count \textit{nRMSE} comparison across hours of the day for validation networks.}
    \begin{tabular}{cccc} 
        \toprule
        \textbf{Network} & \textbf{Baseline} & \textbf{Proposed} & \textbf{\% Improvement} \\
        \midrule
        Los Angeles & 108 &  39 & 64 \\
        \midrule
        San Diego & 110 & 29 & 74 \\
        \bottomrule
    \end{tabular}
    \label{tab:count_nrmse}
\end{table}

\begin{table}[htbp]
    \centering
    \caption{Hour-by-hour count \textit{nRMSE} analysis for validation networks.}
    \label{tab:hourly_count_nrmse}
    \begin{tabular}{lccccr}
        \toprule
        \textbf{Network} & \textbf{Hour of day} & \textbf{Baseline} & \textbf{Proposed} & \textbf{\% Improvement} \\
        \midrule
        \multirow{15}{*}{\textbf{Los Angeles}} & 5 & 109 & 44 & 60 \\
        & 6 & 103 & 65 & 37 \\
        & 7 & 99 & 31 & 69 \\
        & 8 & 98 & 32 & 67 \\
        & 9 & 98 & 31 & 68 \\
        & 10 & 98 & 38 & 61 \\
        & 11 & 98 & 28 & 71 \\
        & 12 & 98 & 27 & 72 \\
        & 13 & 97 & 26 & 73 \\
        & 14 & 96 & 19 & 80 \\
        & 15 & 95 & 28 & 71 \\
        & 16 & 95 & 57 & 40 \\
        & 17 & 95 & 60 & 37 \\
        & 18 & 96 & 51 & 47 \\
        & 19 & 99 & 35 & 65 \\
        \midrule
        \multirow{15}{*}{\textbf{San Diego}} & 5 & 111 & 34 & 69 \\
        & 6 & 105 & 30 & 71 \\
        & 7 & 102 & 50 & 51 \\
        & 8 & 100 & 40 & 60 \\
        & 9 & 99 & 25 & 75 \\
        & 10 & 97 & 21 & 78 \\
        & 11 & 96 & 23 & 76 \\
        & 12 & 96 & 24 & 75 \\
        & 13 & 97 & 20 & 79 \\
        & 14 & 97 & 33 & 66 \\
        & 15 & 98 & 25 & 75 \\ 
        & 16 & 97 & 27 & 72 \\
        & 17 & 97 & 24 & 75 \\
        & 18 & 98 & 37 & 62 \\
        & 19 & 99 & 38 & 62 \\
        \bottomrule
    \end{tabular}
\end{table}

The combination of a low count \textit{nRMSE} and segment count scatter plots exhibiting data points tightly clustered around the identity line shows that the upscaled OD demand using our proposed approach yields physically plausible traffic flow statistics across the network. This outcome serves as robust empirical validation for the analytical scaling approach's efficacy in accurately recovering OD demand derived from subsample ramp-to-ramp count statistics.

\section{Case Studies}
\label{sec:casestudies}
\begin{figure}[htbp]
    \centering
    \begin{subfigure}[b]{0.38\textwidth}
        \centering
        \includegraphics[width=\linewidth]{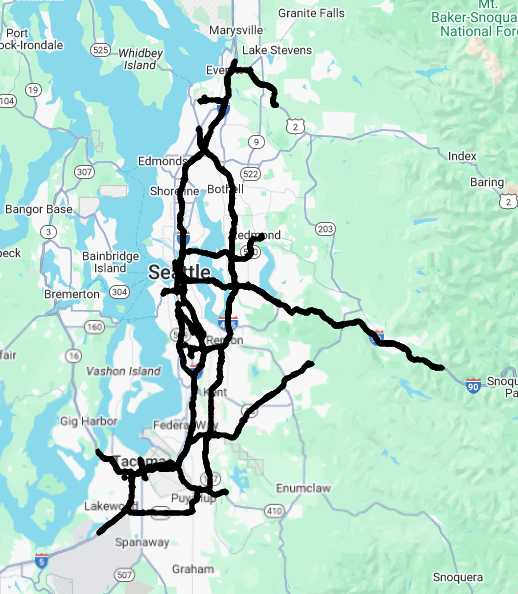}
        \caption{Seattle Highway Network.}
        \label{subfig:seattle_network}
    \end{subfigure}
    
    \begin{subfigure}[b]{0.44\textwidth}
        \centering
        \includegraphics[width=\linewidth]{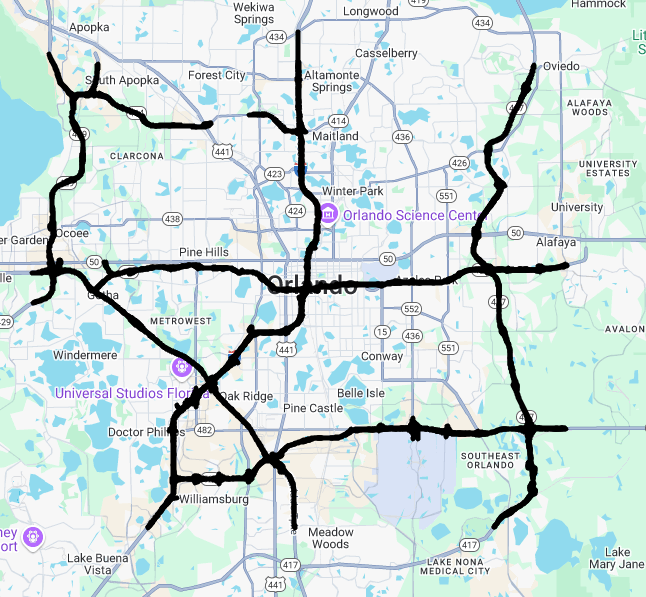}
        \caption{Orlando Highway Network.}
        \label{subfig:orlando_network}
    \end{subfigure}
    \hfill 
    \begin{subfigure}[b]{0.38\textwidth}
        \centering
        \includegraphics[width=\linewidth]{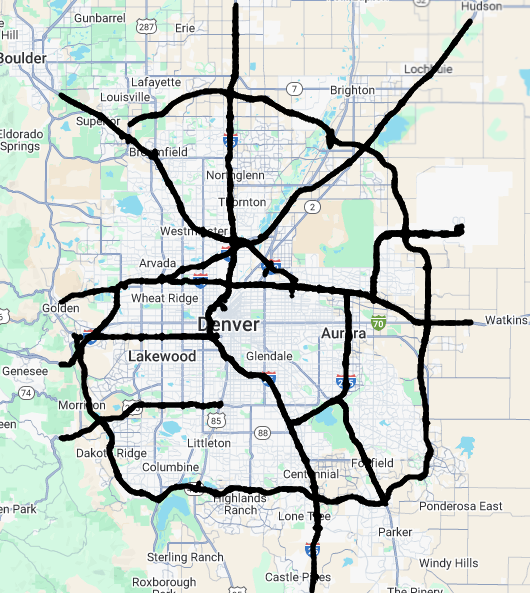}
        \caption{Denver Highway Network.}
        \label{subfig:denver_network}
    \end{subfigure}

    \begin{subfigure}[b]{0.44\textwidth}
        \centering
        \includegraphics[width=\linewidth]{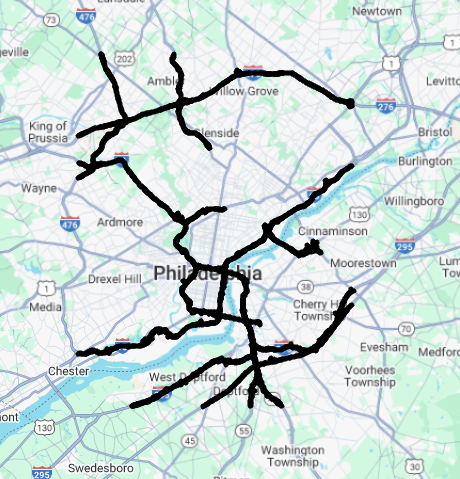}
        \caption{Philadelphia Highway Network.}
        \label{subfig:philly_network}
    \end{subfigure}
    \hfill 
    \begin{subfigure}[b]{0.38\textwidth}
        \centering
        \includegraphics[width=\linewidth]{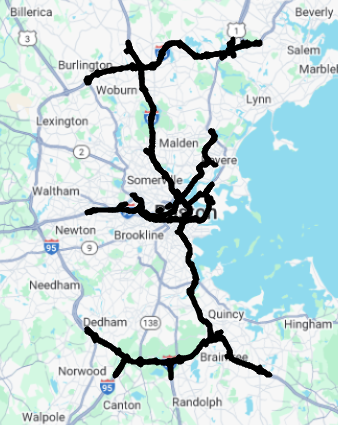}
        \caption{Boston Highway Network.}
        \label{subfig:boston_network}
    \end{subfigure}

    \caption{Case study highway networks. The images are screenshots from Google Maps.}
    \label{fig:case_study_networks}
\end{figure}

This section assesses the scalability of the proposed OD estimation methodology. A critical challenge in OD estimation is the pervasive scarcity of field segment traffic count data. Therefore, our evaluation in this section pivots on travel times as the primary performance metric. Specifically, this refers to ramp-to-ramp route travel times. For the purposes of this discussion, OD travel times and path travel times are considered synonymous and will be referred to as OD travel times throughout this section. The evaluation encompasses a diverse set of five large-scale metropolitan highway networks in the US: Seattle, Denver, Orlando, Philadelphia, and Boston. Their highway networks are presented in Figure \ref{fig:case_study_networks}, with the specifications for the scenario networks provided in Table \ref{tab:case_study_network_spec}.

\begin{table}[htbp]
    \centering
    \caption{Case study network specifications.}
    \begin{tabular}{ccc} 
        \toprule
        \textbf{Network} & \textbf{Number of segments} & \begin{tabular}{@{}c@{}}\textbf{Max hourly}\\\textbf{OD dimensionality}\end{tabular} \\
        \midrule
        Seattle & 18,650 & 451 \\
        \midrule
        Orlando & 8,028 & 417 \\
        \midrule
        Denver & 15,647 & 664 \\
        \midrule
        Philadelphia & 10,709 & 574 \\
        \midrule
        Boston & 6,689 & 312 \\
        \bottomrule
    \end{tabular}
    \label{tab:case_study_network_spec}
\end{table}

To evaluate the quality of the proposed ODs, we use SUMO \citep{Sumo18}, a high-resolution traffic simulation model for each network. To evaluate the performance of a given OD, we simulate its performance and compute the \textit{nRMSE} of the OD travel times. This \textit{nRMSE} is computed by comparing the simulated travel times with the ground truth (i.e., field data) travel times. The use of the simulator allows us to evaluate the quality of an OD with a model that was not used for estimation purposes and that is expected to be of higher fidelity. 

For each network and each time interval of interest, we compare the performance of three ODs: (1) the baseline OD, which consists of the original unscaled subsample OD; (2) the proposed OD; and (3) the benchmark OD. The benchmark OD is obtained as follows. We do a fine-grained grid-search over all possible scaling factor values ($x$ values), each upscaled OD is simulated and the OD with the lowest travel time \textit{nRMSE} is selected as the benchmark OD. 

Note that since the simulator is used to evaluate the OD travel time \textit{nRMSE}, the \textit{nRMSE} of the benchmark is a lower bound for that of the proposed method. Hence, the comparison of the \textit{nRMSE} of the proposed method with that of the benchmark serves to illustrate the loss in accuracy due to using a simplified analytical network model to determine the scaling factor rather than using a high-resolution simulator. In other words, this benchmark serves to quantify the potential loss of quality incurred by employing a simpler analytical network model during the OD estimation process. The closer the \textit{nRMSE} of our proposed method approaches that of the benchmark, the higher its demonstrated fidelity in estimating network travel times.

\begin{figure}[htbp]
    \centering
    \begin{subfigure}[b]{0.3\textwidth}
        \centering
        \includegraphics[width=\linewidth]{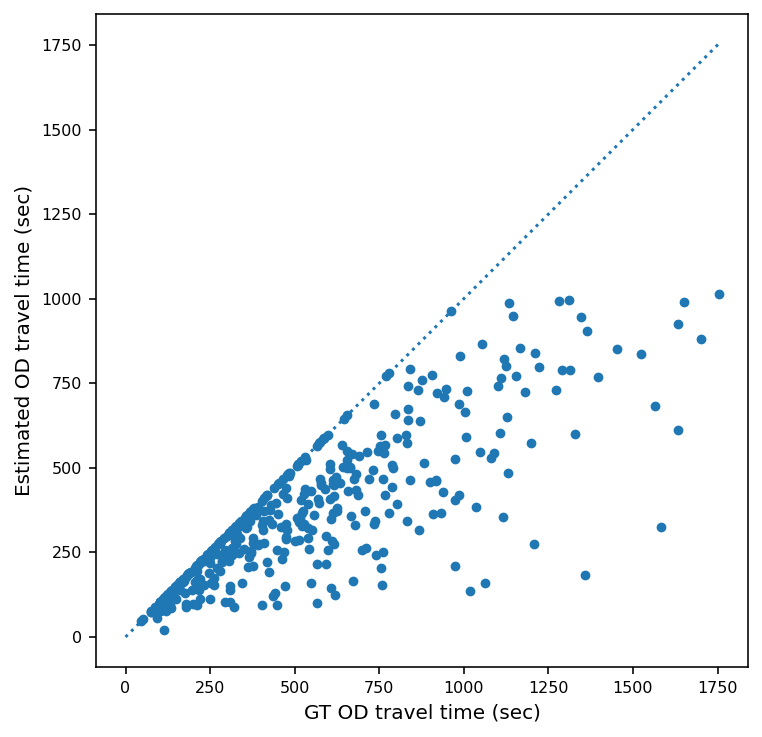}
        \caption{Seattle: Hour 13 Baseline.}
        \label{subfig:seattle_eta_scatter_1pm_baseline}
    \end{subfigure}
    \hfill 
    \begin{subfigure}[b]{0.3\textwidth}
        \centering
        \includegraphics[width=\linewidth]{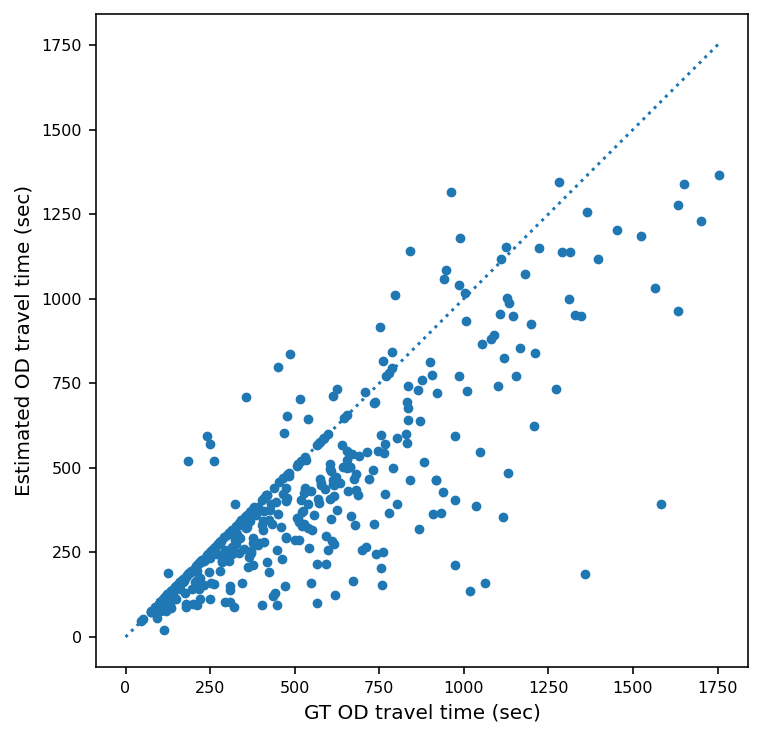}
        \caption{Seattle: Hour 13 Benchmark.}
        \label{subfig:seattle_eta_scatter_1pm_benchmark}
    \end{subfigure}
    \hfill 
    \begin{subfigure}[b]{0.3\textwidth}
        \centering
        \includegraphics[width=\linewidth]{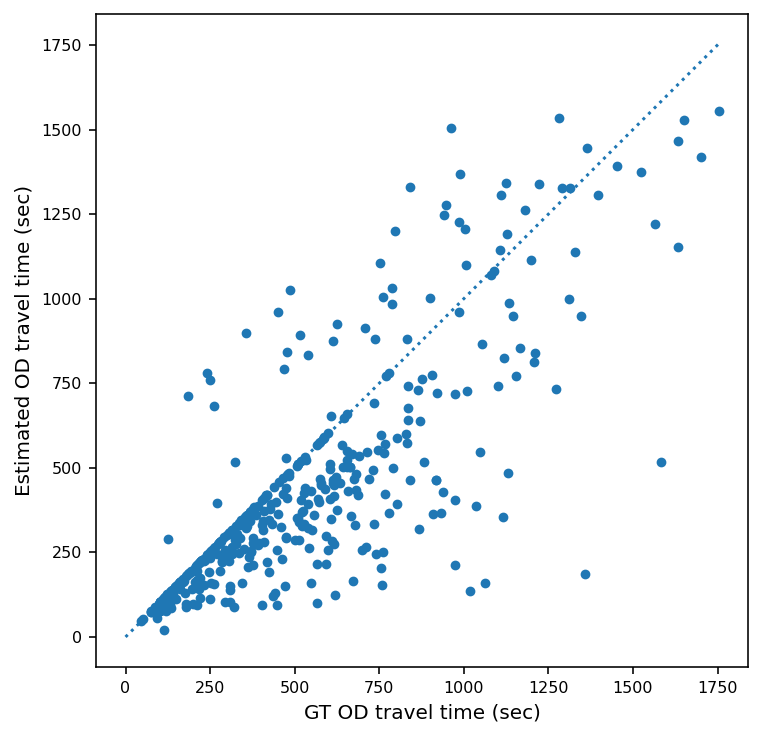}
        \caption{Seattle: Hour 13 Proposed.}
        \label{subfig:seattle_eta_scatter_1pm_proposed}
    \end{subfigure}
    
    \begin{subfigure}[b]{0.3\textwidth}
        \centering
        \includegraphics[width=\linewidth]{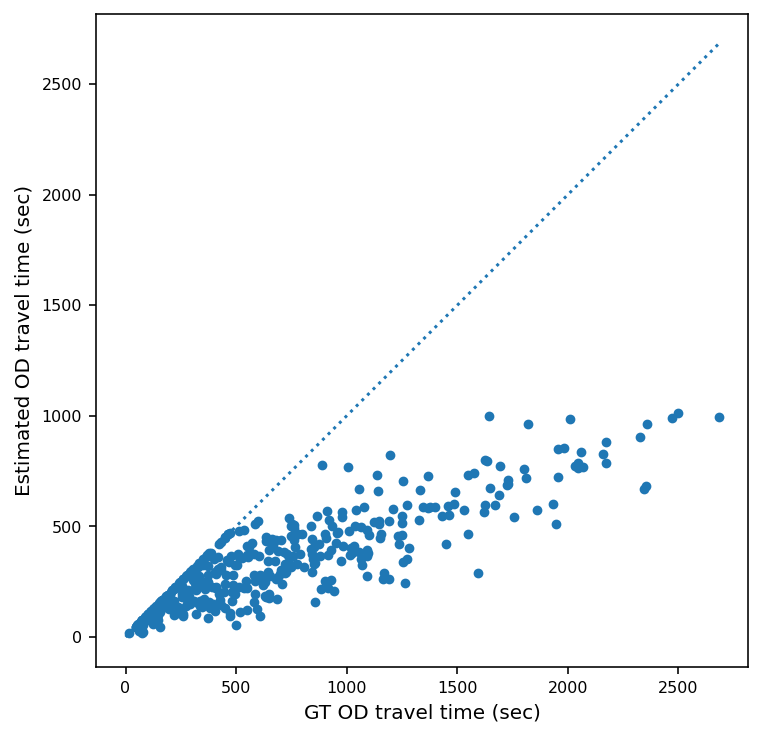}
        \caption{Seattle: Hour 16 Baseline.}
        \label{subfig:seattle_eta_scatter_4pm_baseline}
    \end{subfigure}
    \hfill 
    \begin{subfigure}[b]{0.3\textwidth}
        \centering
        \includegraphics[width=\linewidth]{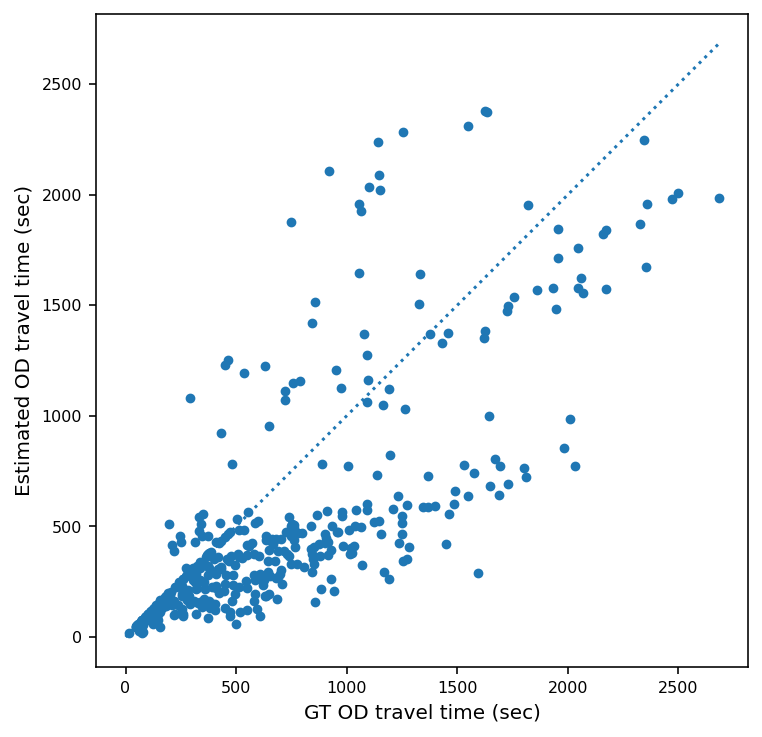}
        \caption{Seattle: Hour 16 Benchmark.}
        \label{subfig:seattle_eta_scatter_4pm_benchmark}
    \end{subfigure}
    \hfill 
    \begin{subfigure}[b]{0.3\textwidth}
        \centering
        \includegraphics[width=\linewidth]{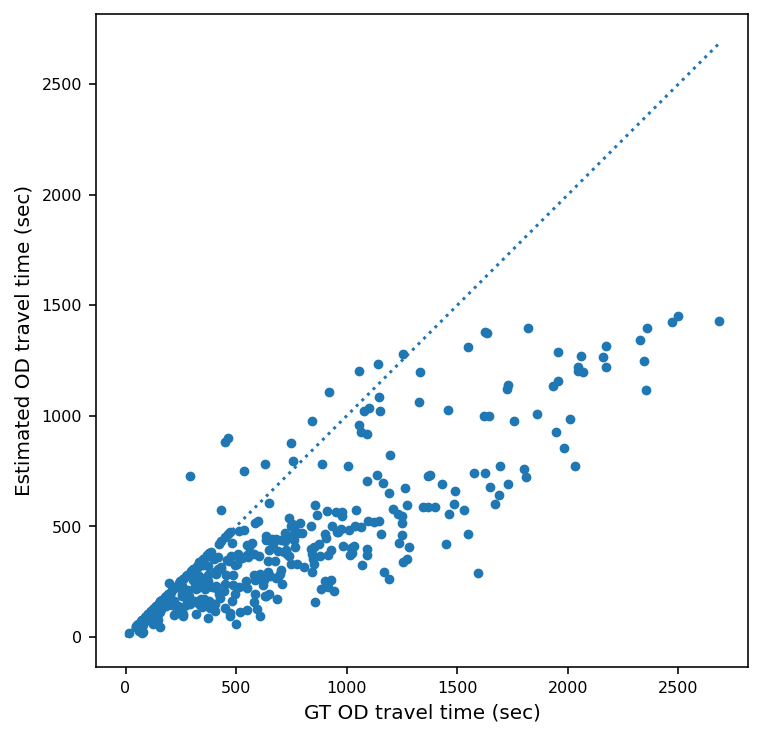}
        \caption{Seattle: Hour 16 Proposed.}
        \label{subfig:seattle_eta_scatter_4pm_proposed}
    \end{subfigure}

    \begin{subfigure}[b]{0.3\textwidth}
        \centering
        \includegraphics[width=\linewidth]{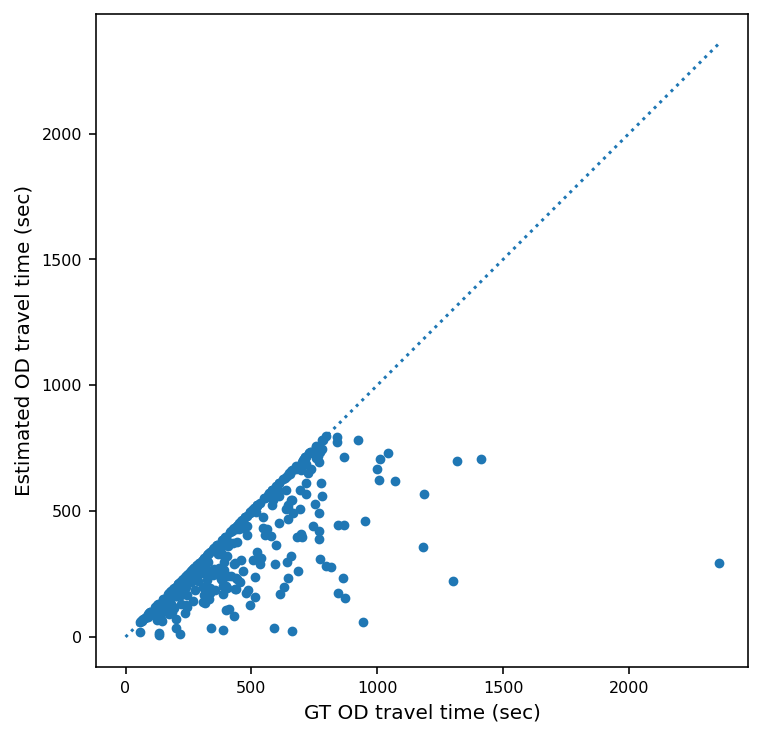}
        \caption{Orlando: Hour 13 Baseline.}
        \label{subfig:orlando_eta_scatter_1pm_baseline}
    \end{subfigure}
    \hfill 
    \begin{subfigure}[b]{0.3\textwidth}
        \centering
        \includegraphics[width=\linewidth]{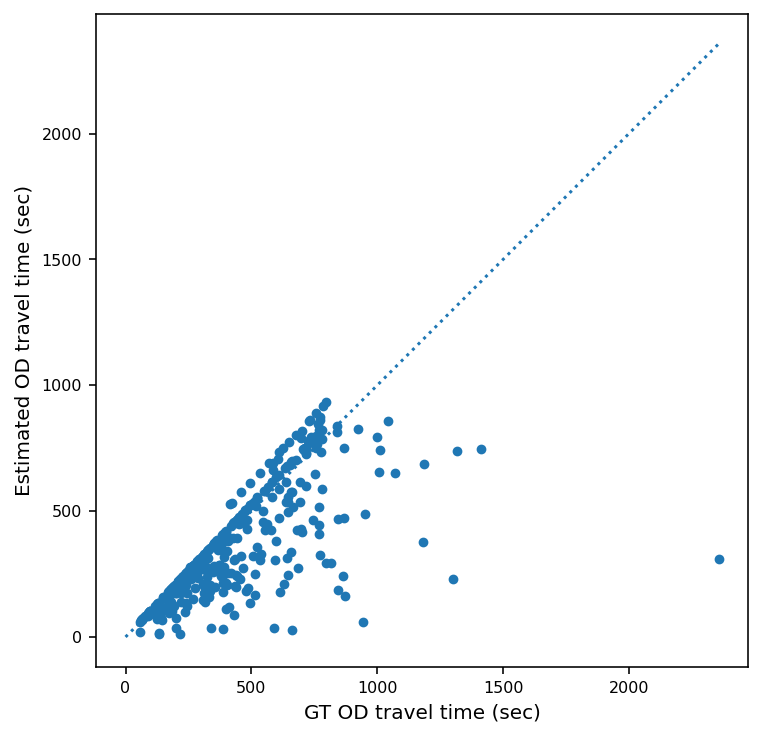}
        \caption{Orlando: Hour 13 Benchmark.}
        \label{subfig:orlando_eta_scatter_1pm_benchmark}
    \end{subfigure}
    \hfill 
    \begin{subfigure}[b]{0.3\textwidth}
        \centering
        \includegraphics[width=\linewidth]{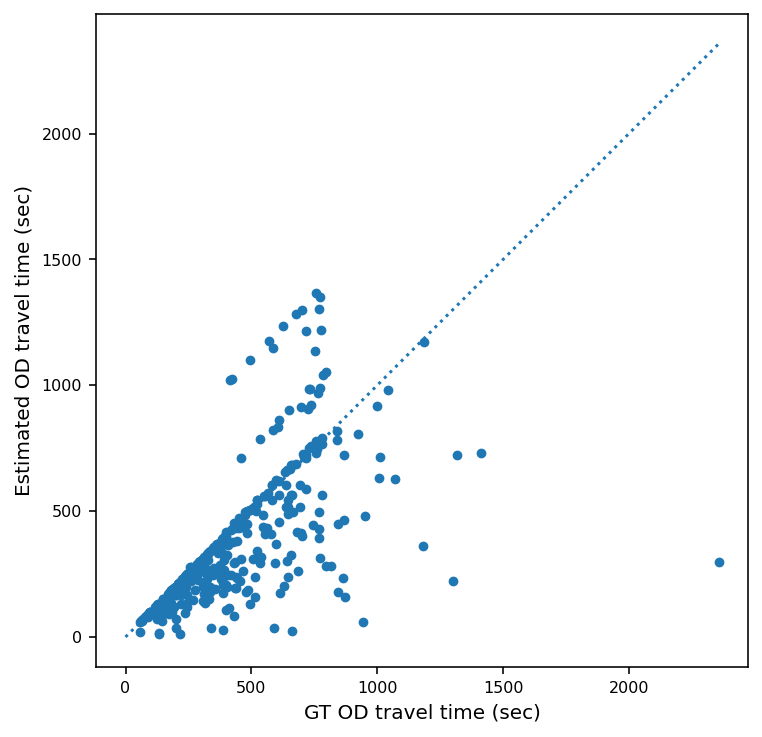}
        \caption{Orlando: Hour 13 Proposed.}
        \label{subfig:orlando_eta_scatter_1pm_proposed}
    \end{subfigure}
    
    \begin{subfigure}[b]{0.3\textwidth}
        \centering
        \includegraphics[width=\linewidth]{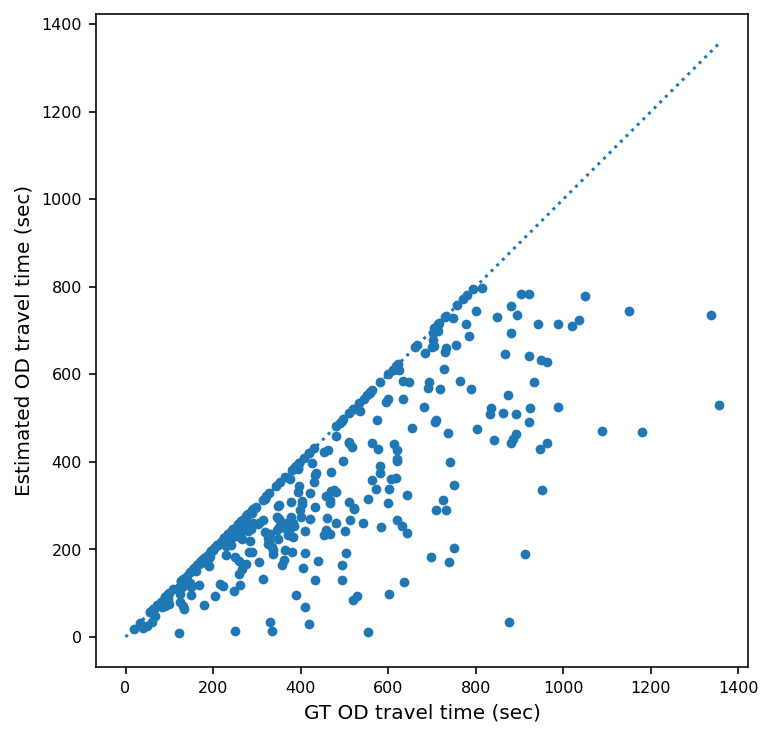}
        \caption{Orlando: Hour 16 Baseline.}
        \label{subfig:orlando_eta_scatter_4pm_baseline}
    \end{subfigure}
    \hfill 
    \begin{subfigure}[b]{0.3\textwidth}
        \centering
        \includegraphics[width=\linewidth]{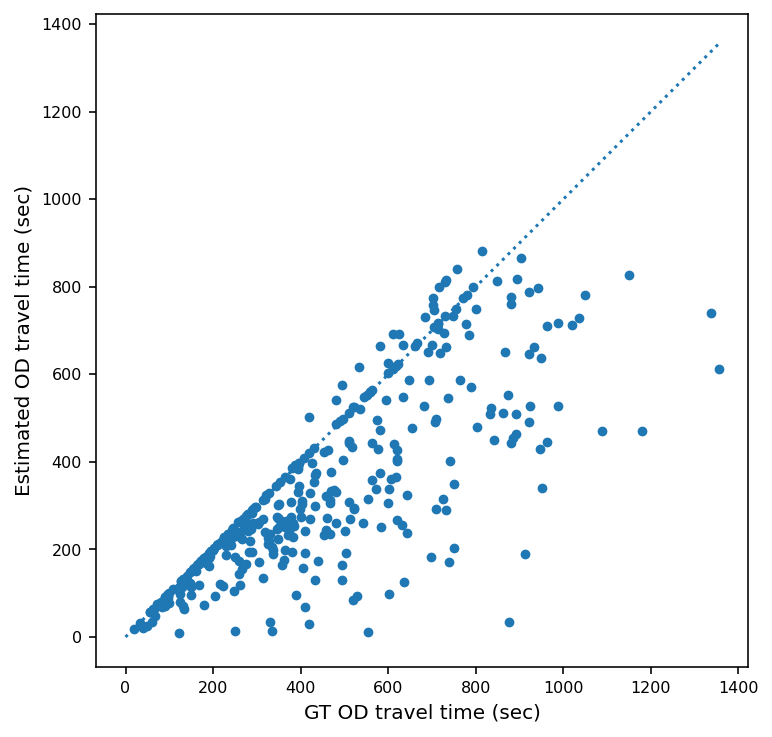}
        \caption{Orlando: Hour 16 Benchmark.}
        \label{subfig:orlando_eta_scatter_4pm_benchmark}
    \end{subfigure}
    \hfill 
    \begin{subfigure}[b]{0.3\textwidth}
        \centering
        \includegraphics[width=\linewidth]{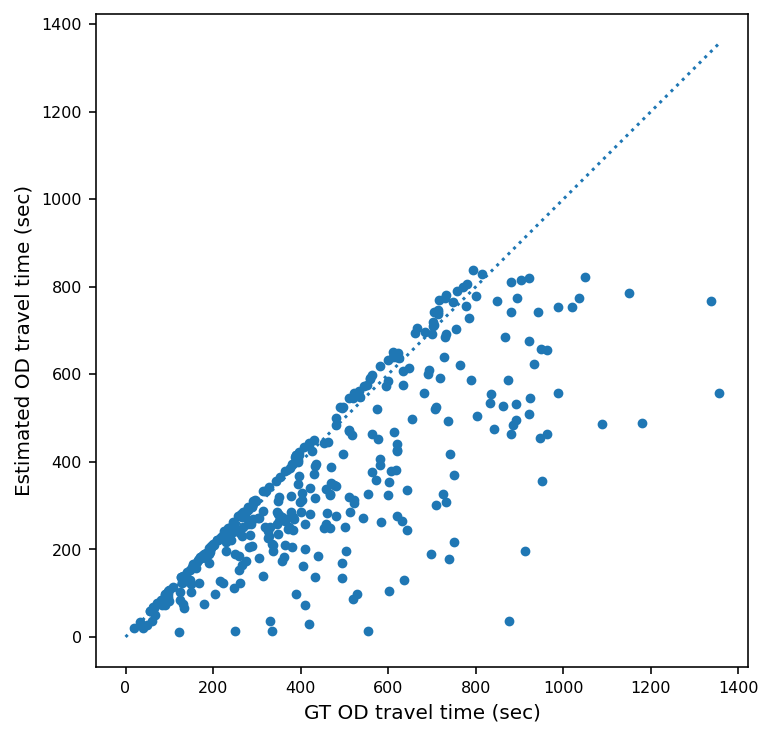}
        \caption{Orlando: Hour 16 Proposed.}
        \label{subfig:orlando_eta_scatter_4pm_proposed}
    \end{subfigure}

    \caption{Simulated OD travel time quality analysis for case study networks - Scatter plot.}
    \label{fig:sim_eta_scatter_part1}
\end{figure}

\begin{figure}[htbp]
    \centering
    \begin{subfigure}[b]{0.3\textwidth}
        \centering
        \includegraphics[width=\linewidth]{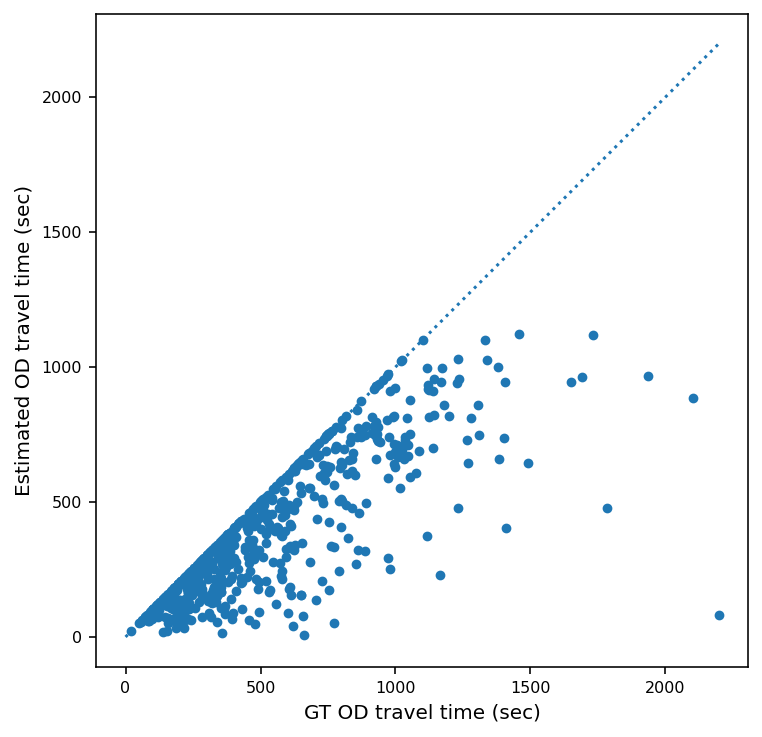}
        \caption{Denver: Hour 13 Baseline.}
        \label{subfig:denver_eta_scatter_1pm_baseline}
    \end{subfigure}
    \hfill 
    \begin{subfigure}[b]{0.3\textwidth}
        \centering
        \includegraphics[width=\linewidth]{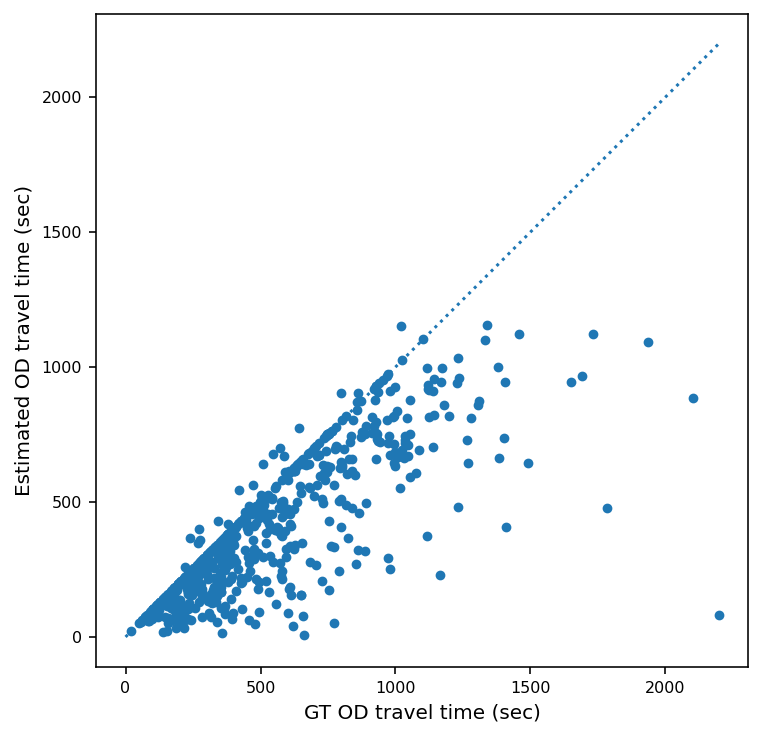}
        \caption{Denver: Hour 13 Benchmark.}
        \label{subfig:denver_eta_scatter_1pm_benchmark}
    \end{subfigure}
    \hfill 
    \begin{subfigure}[b]{0.3\textwidth}
        \centering
        \includegraphics[width=\linewidth]{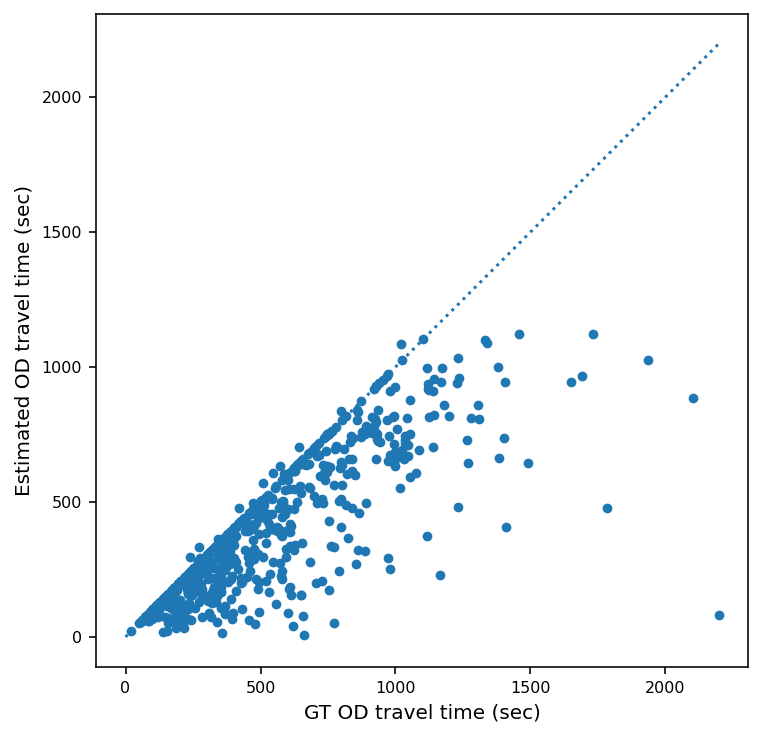}
        \caption{Denver: Hour 13 Proposed.}
        \label{subfig:denver_eta_scatter_1pm_proposed}
    \end{subfigure}
    
    \begin{subfigure}[b]{0.3\textwidth}
        \centering
        \includegraphics[width=\linewidth]{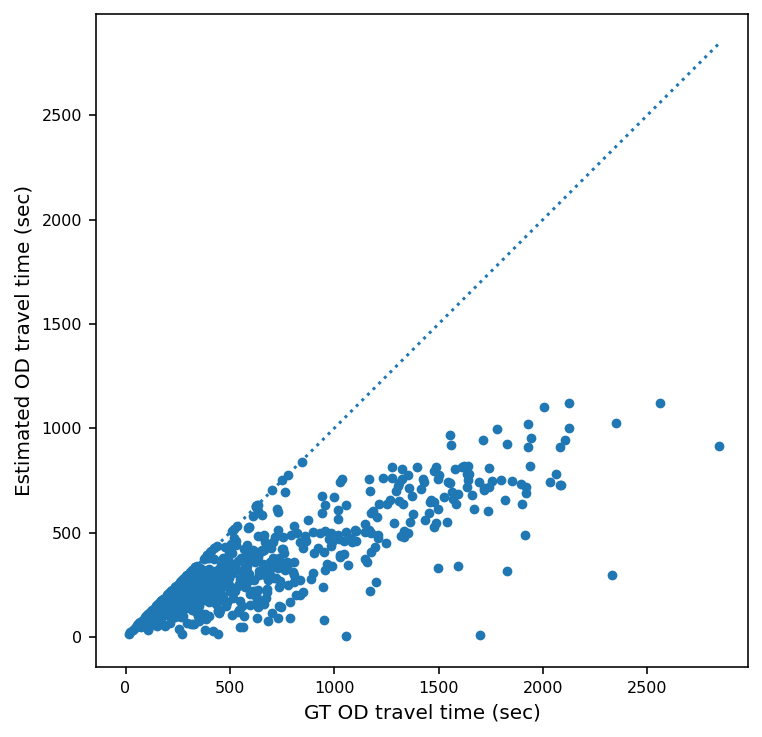}
        \caption{Denver: Hour 16 Baseline.}
        \label{subfig:denver_eta_scatter_4pm_baseline}
    \end{subfigure}
    \hfill 
    \begin{subfigure}[b]{0.3\textwidth}
        \centering
        \includegraphics[width=\linewidth]{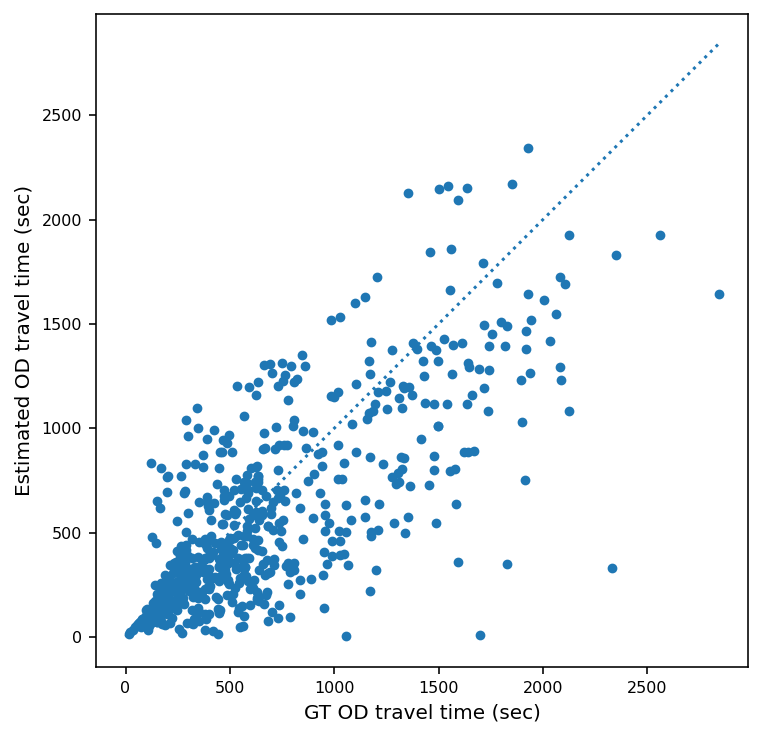}
        \caption{Denver: Hour 16 Benchmark.}
        \label{subfig:denver_eta_scatter_4pm_benchmark}
    \end{subfigure}
    \hfill 
    \begin{subfigure}[b]{0.3\textwidth}
        \centering
        \includegraphics[width=\linewidth]{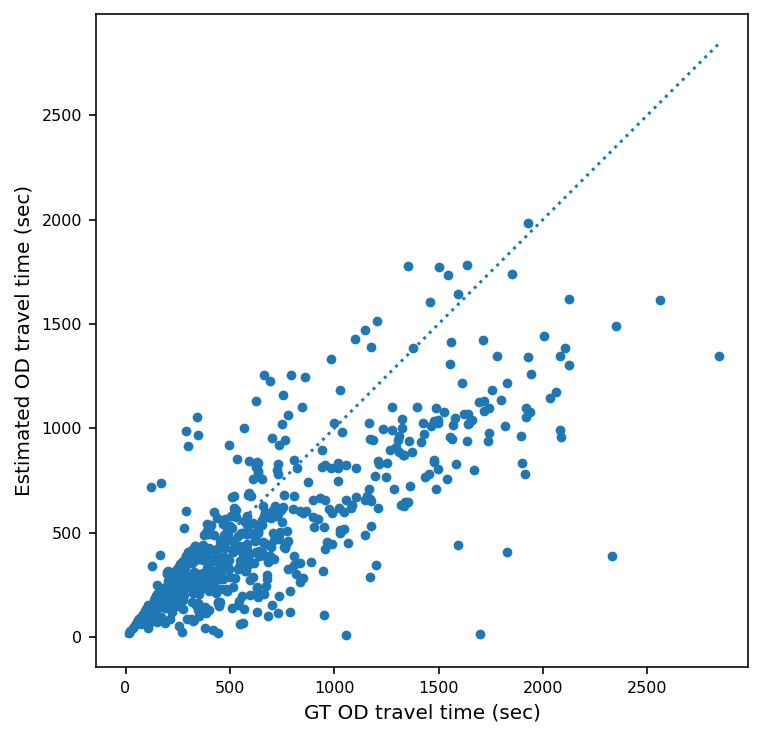}
        \caption{Denver: Hour 16 Proposed.}
        \label{subfig:denver_eta_scatter_4pm_proposed}
    \end{subfigure}
    
    \begin{subfigure}[b]{0.3\textwidth}
        \centering
        \includegraphics[width=\linewidth]{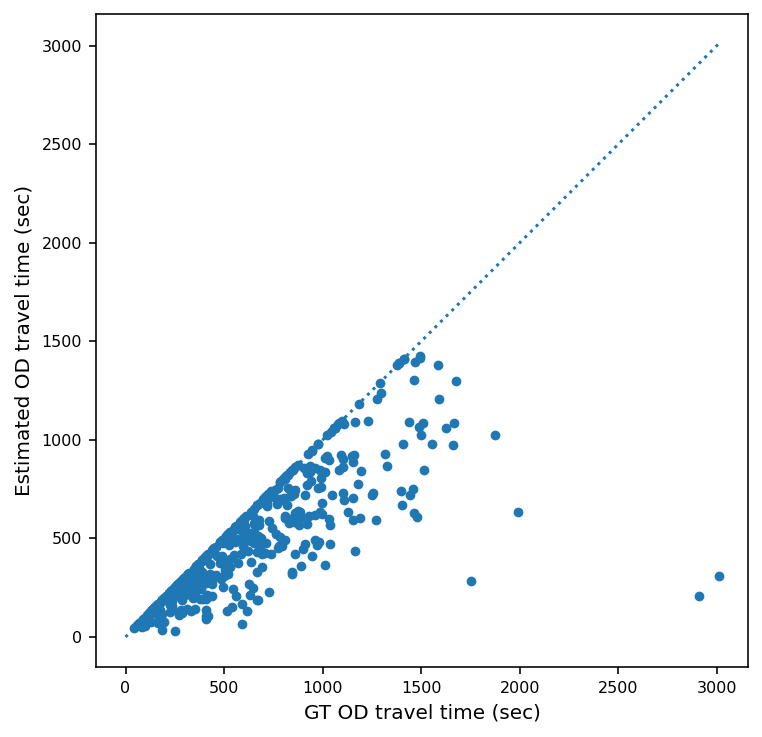}
        \caption{Philadelphia: Hour 13 Baseline.}
        \label{subfig:philly_eta_scatter_1pm_baseline}
    \end{subfigure}
    \hfill 
    \begin{subfigure}[b]{0.3\textwidth}
        \centering
        \includegraphics[width=\linewidth]{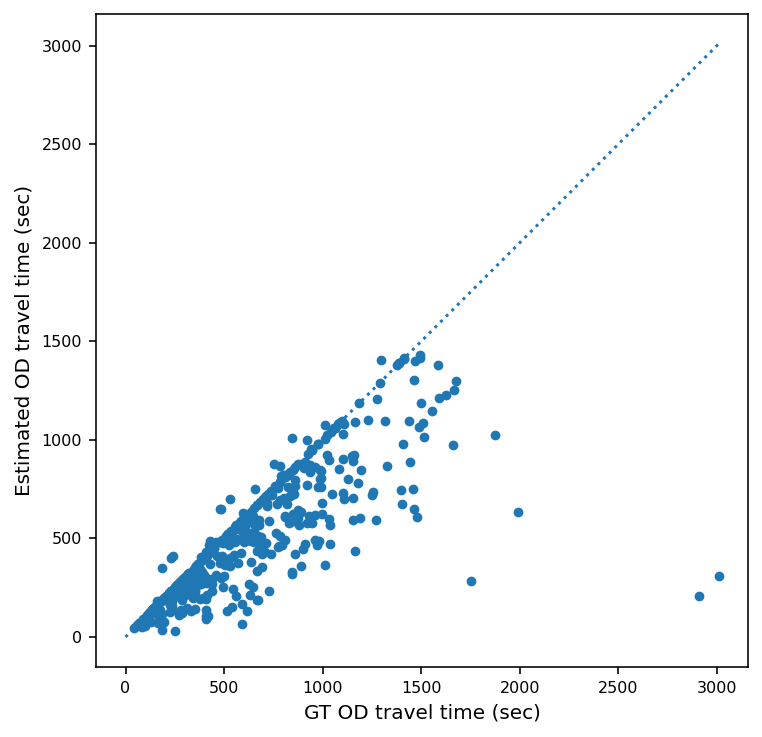}
        \caption{Philadelphia: Hour 13 Benchmark.}
        \label{subfig:philly_eta_scatter_1pm_benchmark}
    \end{subfigure}
    \hfill 
    \begin{subfigure}[b]{0.3\textwidth}
        \centering
        \includegraphics[width=\linewidth]{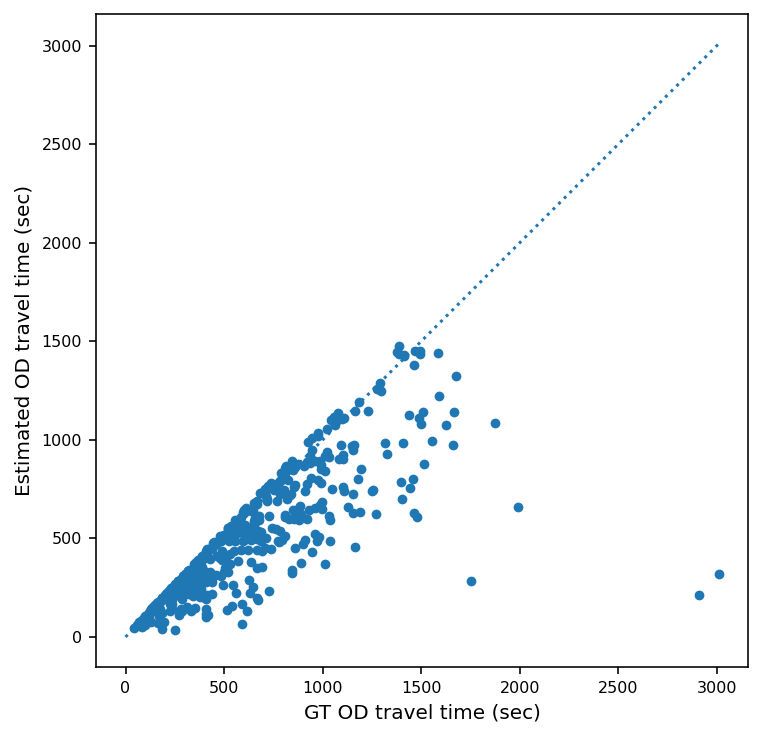}
        \caption{Philadelphia: Hour 13 Proposed.}
        \label{subfig:philly_eta_scatter_1pm_proposed}
    \end{subfigure}
    
    \begin{subfigure}[b]{0.3\textwidth}
        \centering
        \includegraphics[width=\linewidth]{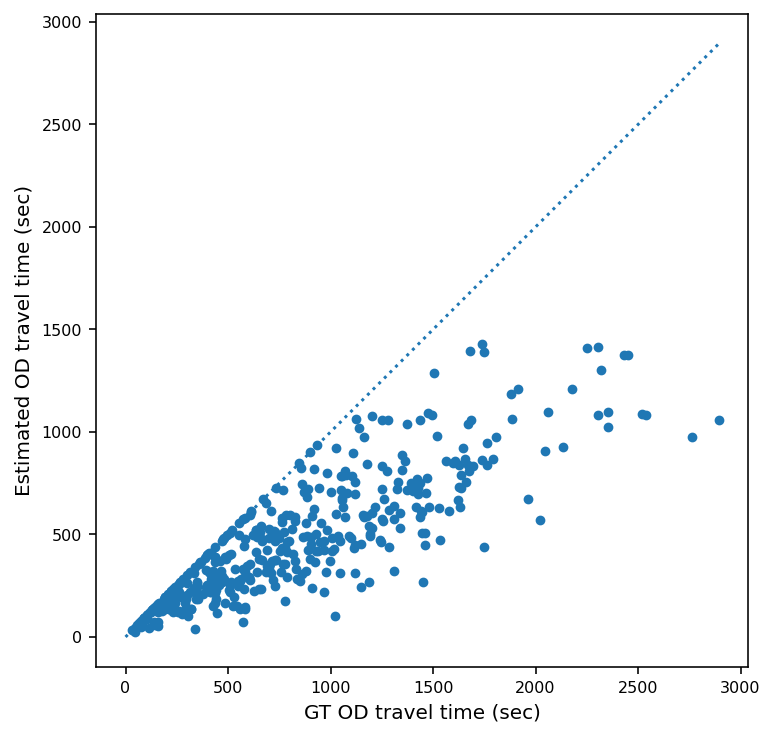}
        \caption{Philadelphia: Hour 16 Baseline.}
        \label{subfig:philly_eta_scatter_4pm_baseline}
    \end{subfigure}
    \hfill 
    \begin{subfigure}[b]{0.3\textwidth}
        \centering
        \includegraphics[width=\linewidth]{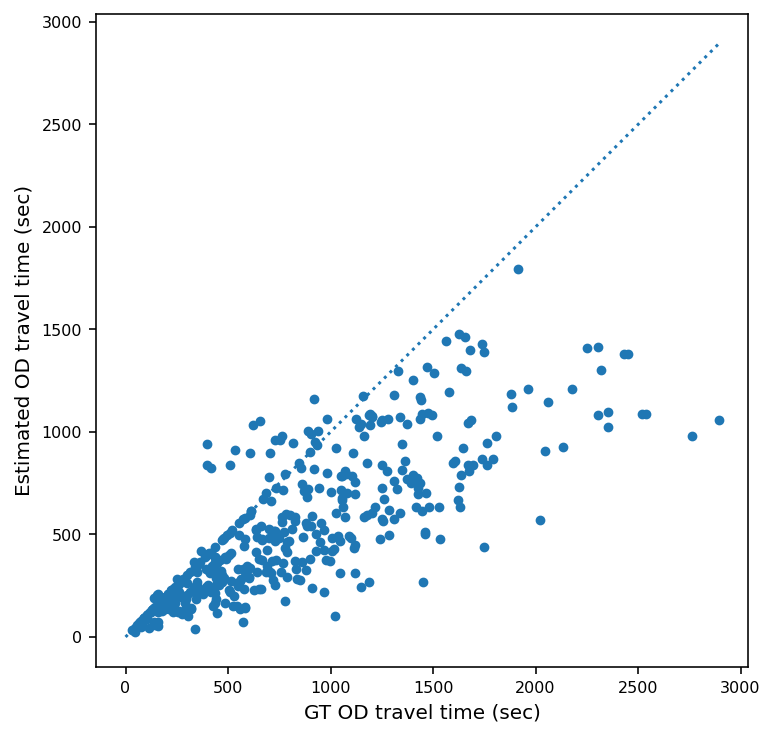}
        \caption{Philadelphia: Hour 16 Benchmark.}
        \label{subfig:philly_eta_scatter_4pm_benchmark}
    \end{subfigure}
    \hfill 
    \begin{subfigure}[b]{0.3\textwidth}
        \centering
        \includegraphics[width=\linewidth]{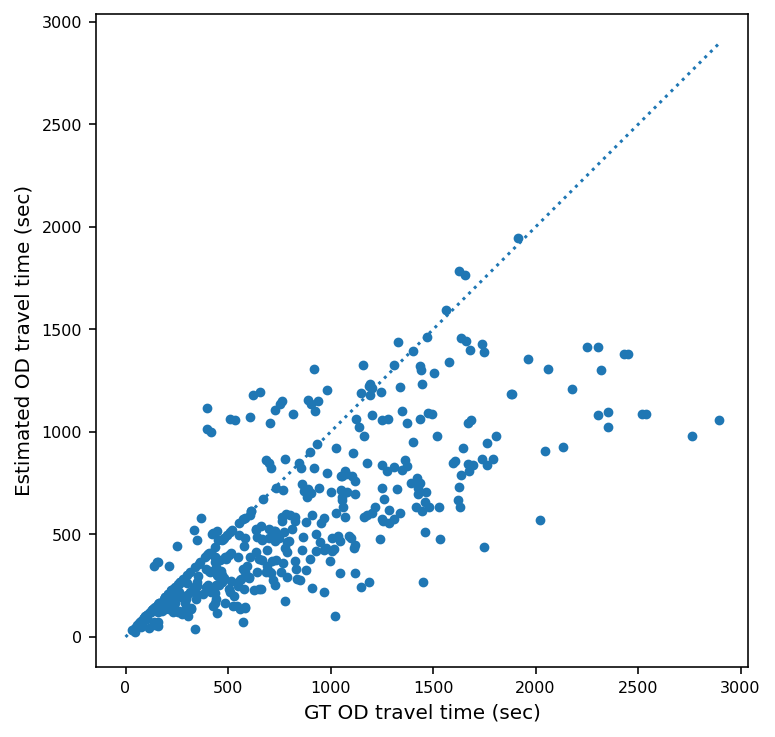}
        \caption{Philadelphia: Hour 16 Proposed.}
        \label{subfig:philly_eta_scatter_4pm_proposed}
    \end{subfigure}

    \caption{Simulated OD travel time quality analysis for case study networks - Scatter plot (continued).}
    \label{fig:sim_eta_scatter_part2}
\end{figure}

\begin{figure}[htbp]
    \centering
    \begin{subfigure}[b]{0.3\textwidth}
        \centering
        \includegraphics[width=\linewidth]{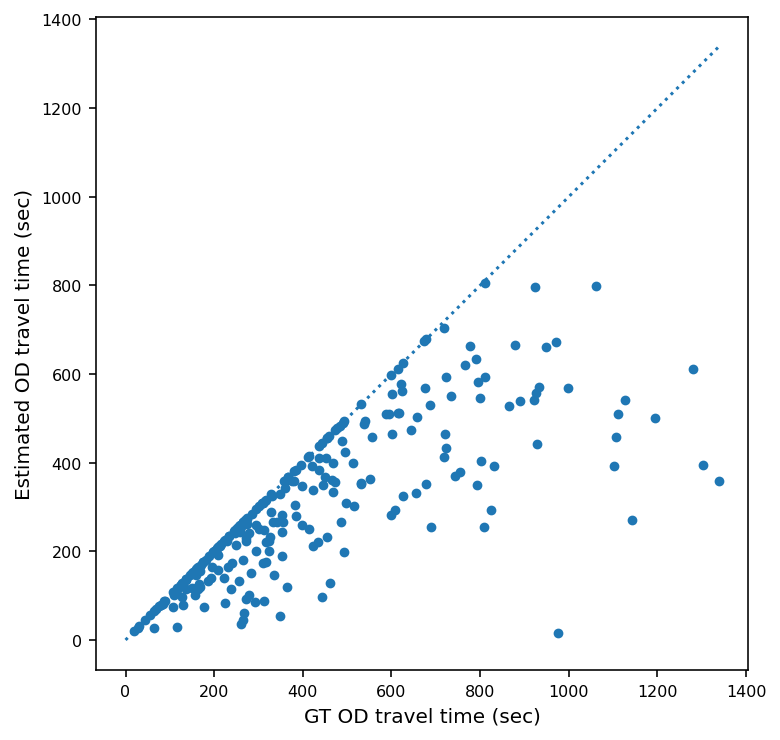}
        \caption{Boston: Hour 13 Baseline.}
        \label{subfig:boston_eta_scatter_1pm_baseline}
    \end{subfigure}
    \hfill 
    \begin{subfigure}[b]{0.3\textwidth}
        \centering
        \includegraphics[width=\linewidth]{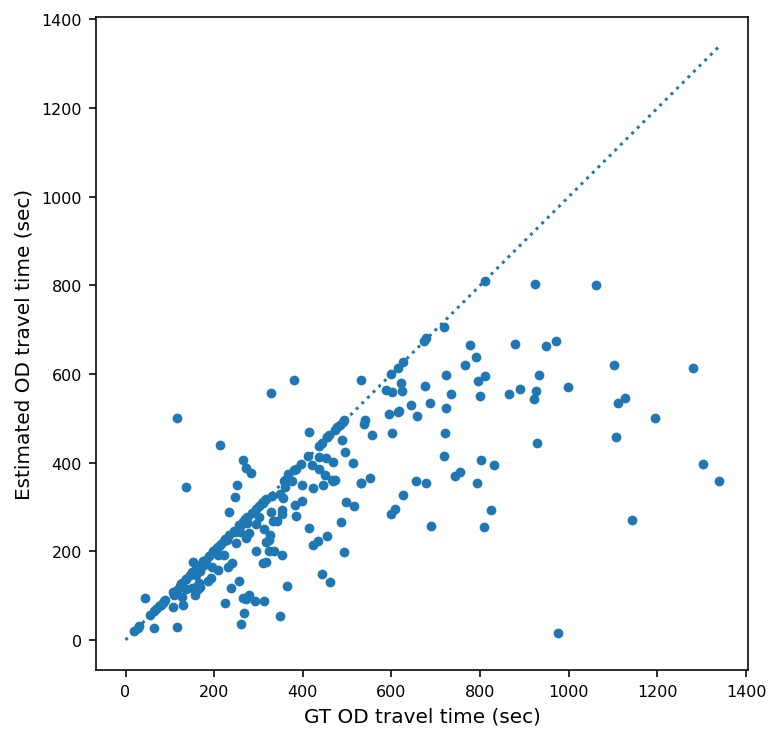}
        \caption{Boston: Hour 13 Benchmark.}
        \label{subfig:boston_eta_scatter_1pm_benchmark}
    \end{subfigure}
    \hfill 
    \begin{subfigure}[b]{0.3\textwidth}
        \centering
        \includegraphics[width=\linewidth]{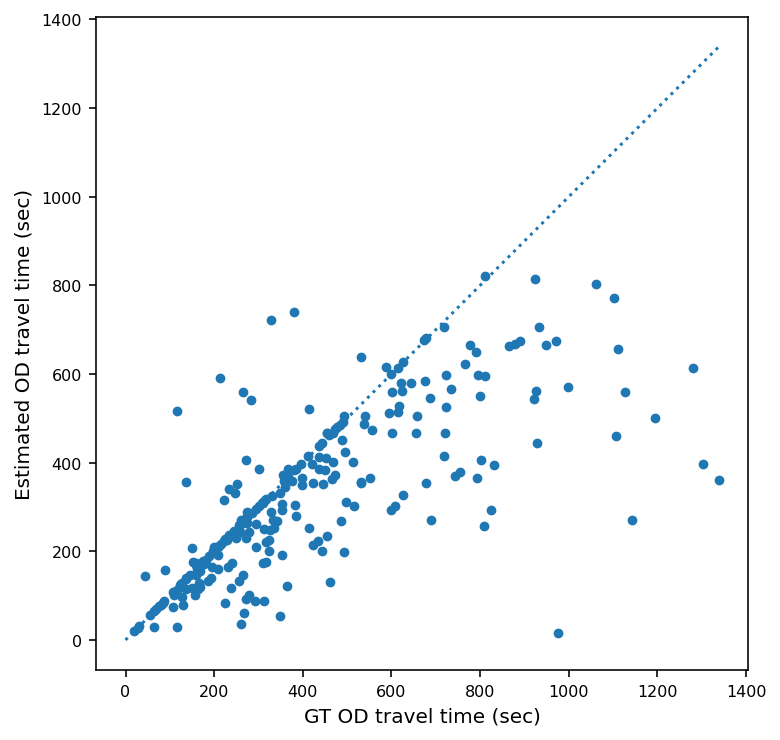}
        \caption{Boston: Hour 13 Proposed.}
        \label{subfig:boston_eta_scatter_1pm_proposed}
    \end{subfigure}
    
    \begin{subfigure}[b]{0.3\textwidth}
        \centering
        \includegraphics[width=\linewidth]{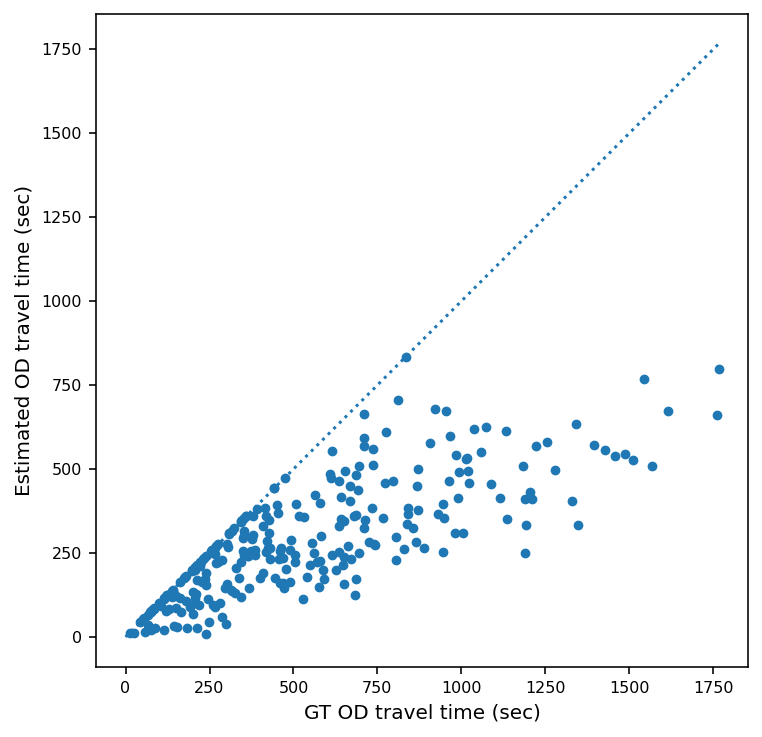}
        \caption{Boston: Hour 16 Baseline.}
        \label{subfig:boston_eta_scatter_4pm_baseline}
    \end{subfigure}
    \hfill 
    \begin{subfigure}[b]{0.3\textwidth}
        \centering
        \includegraphics[width=\linewidth]{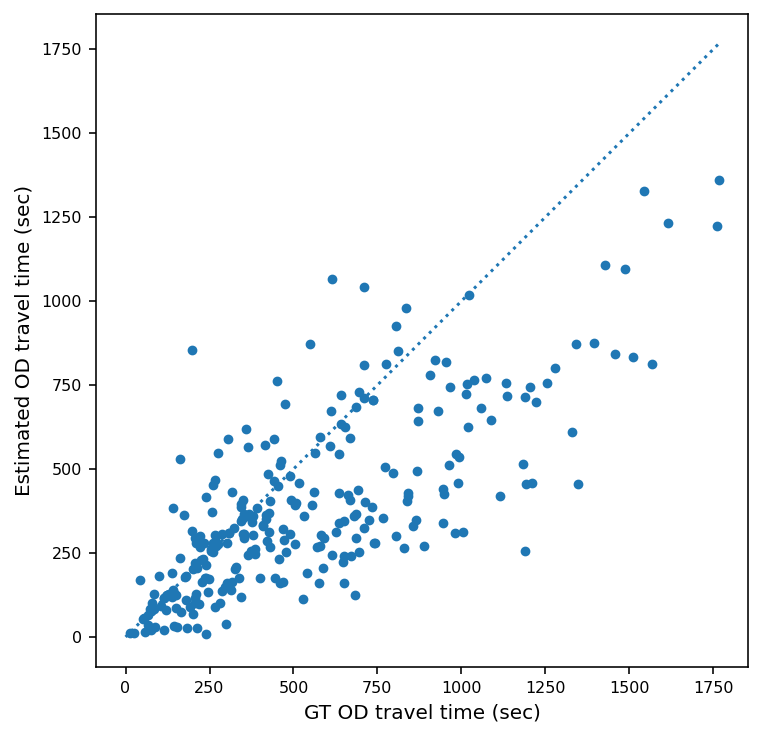}
        \caption{Boston: Hour 16 Benchmark.}
        \label{subfig:boston_eta_scatter_4pm_benchmark}
    \end{subfigure}
    \hfill 
    \begin{subfigure}[b]{0.3\textwidth}
        \centering
        \includegraphics[width=\linewidth]{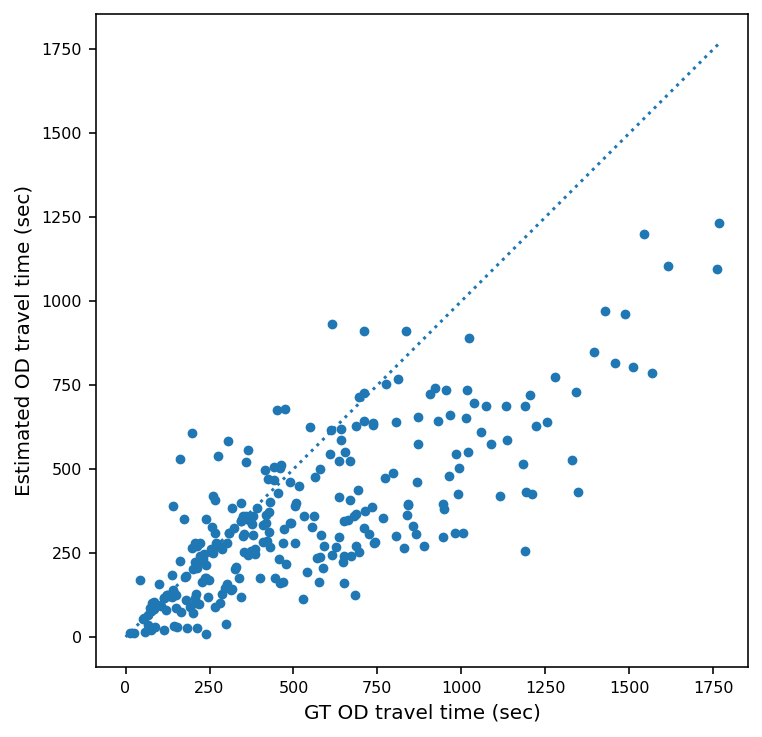}
        \caption{Boston: Hour 16 Proposed.}
        \label{subfig:boston_eta_scatter_4pm_proposed}
    \end{subfigure}
    
    \caption{Simulated OD travel time quality analysis for case study networks - Scatter plot (continued).}
    \label{fig:sim_eta_scatter_part3}
\end{figure}

Figures \ref{fig:sim_eta_scatter_part1}-\ref{fig:sim_eta_scatter_part3} present scatter plots comparing ground truth (\textit{x}-axis) and estimated (\textit{y}-axis) OD travel times for five case study cities and for each hour. For every city and hour, we provide a visual comparison of three distinct methods: baseline, benchmark, and proposed (arranged from left to right in the subplots). To ensure direct comparability, all three plots in a row share the same \textit{x} and \textit{y}-axes ranges. The proximity of data points to the identity line directly reflects the quality of estimated OD travel times of a given method. We analyze two scenarios per city to represent different congestion levels: hour 13 (1-2 PM), a typical afternoon off-peak period, and hour 16 (4-5 PM), a typical afternoon peak period. All times are reported in the respective city's local time. These two scenarios are displayed as separate rows of subplots for each city. For example, Seattle's off-peak performance is shown in Subplots \ref{subfig:seattle_eta_scatter_1pm_baseline}-\ref{subfig:seattle_eta_scatter_1pm_proposed}, while its peak performance is presented in Subplots \ref{subfig:seattle_eta_scatter_4pm_baseline}-\ref{subfig:seattle_eta_scatter_4pm_proposed}. A general trend observed is that congestion severity is noticeably higher during afternoon peak hours compared to off-peak hours. This is evident in the expanded range of both the \textit{x} and \textit{y}-axes in the scatter plots for hour 16 scenarios, indicating longer path travel times. However, the extent of this phenomenon varies across cities. For instance, Orlando shows less distinct differences between peak and off-peak periods, with overall mild congestion.

Across all cities, the baseline (left column of subplots) demonstrates a better approximation to ground truth during off-peak scenarios than during peak scenarios. This suggests that the baseline performs more effectively in replicating off-peak travel times. The baseline's considerably good approximation to ground truth for many cities in off-peak conditions is anticipated, as traffic during these periods is predominantly free-flowing and can be accurately replicated when demand is relatively low. For off-peak conditions, our proposed approach generally aligns with the benchmark method. However, for cities like Denver (Subplots \ref{subfig:denver_eta_scatter_1pm_baseline}-\ref{subfig:denver_eta_scatter_1pm_proposed}) and Philadelphia (Subplots \ref{subfig:philly_eta_scatter_1pm_baseline}-\ref{subfig:philly_eta_scatter_1pm_proposed}), neither the benchmark nor the proposed method shows significant improvement over the baseline, with very similar data patterns observed across all three methods. In contrast, a visual distinction is more apparent for Seattle (Subplots \ref{subfig:seattle_eta_scatter_1pm_baseline}-\ref{subfig:seattle_eta_scatter_1pm_proposed}), Orlando (Subplots \ref{subfig:orlando_eta_scatter_1pm_baseline}-\ref{subfig:orlando_eta_scatter_1pm_proposed}), and Boston (Subplots \ref{subfig:boston_eta_scatter_1pm_baseline}-\ref{subfig:boston_eta_scatter_1pm_proposed}), where both the benchmark and proposed methods exhibit an upshifting trend. This results in data appearing more evenly distributed around the identity line, a trend most evident in Seattle. 

Significant distinctions in performance across methods are observed during peak hour scenarios. The baseline method consistently exhibits underestimation during these periods, particularly in Seattle (Subplots \ref{subfig:seattle_eta_scatter_4pm_baseline}-\ref{subfig:seattle_eta_scatter_4pm_proposed}), Denver (Subplots \ref{subfig:denver_eta_scatter_4pm_baseline}-\ref{subfig:denver_eta_scatter_4pm_proposed}), Philadelphia (Subplots \ref{subfig:philly_eta_scatter_4pm_baseline}-\ref{subfig:philly_eta_scatter_4pm_proposed}), and Boston (Subplots \ref{subfig:boston_eta_scatter_4pm_baseline}-\ref{subfig:boston_eta_scatter_4pm_proposed}). For these cities, nearly all baseline data points fall below the identity line, with a substantial number being considerably distant from it. This pronounced underestimation underscores the critical importance of subsample demand upscaling especially in peak demand situations. Overall, the benchmark method demonstrates effective upscaling, successfully shifting the data cluster to align closely with the identity line. This highlights the benchmark's capability in accurately capturing peak hour conditions compared to the baseline. Our proposed method shows strong agreement with the benchmark in Orlando (Subplots \ref{subfig:orlando_eta_scatter_4pm_baseline}-\ref{subfig:orlando_eta_scatter_4pm_proposed}), Denver, Philadelphia, and Boston. However, in Seattle, the magnitude of upscaling achieved by the proposed method lies between that of the baseline and the benchmark. This indicates a milder upscaling effect compared to the benchmark in Seattle, signifying a quality gap between our proposed model and the high-resolution simulator in this specific context.

Table \ref{tab:hourly_eta_nrmse_simulated} presents a quantitative assessment of \textit{nRMSE} of simulated OD travel times across an expanded set of hours. This analysis extends beyond the previously discussed scenarios in Figures \ref{fig:sim_eta_scatter_part1}-\ref{fig:sim_eta_scatter_part3} by encompassing five distinct hours, ranging from 1 PM to 6 PM, with a separate evaluation conducted for each individual hour. Across the five hours analyzed for each of the five cities, the period from 4 PM to 6 PM generally exhibits a higher level of congestion compared to the 1 PM to 3 PM period. Hereafter, we will refer to the 4-6 PM timeframe as peak hours and the 1-3 PM timeframe as off-peak hours. In every instance, our proposed method's \textit{nRMSE} is directly compared against both the baseline and the benchmark. All \textit{nRMSE} are computed based on simulated travel times and are presented as percentages, with the percentage symbol omitted for conciseness. Additionally, the \%Gap metric, presented in the rightmost column of the table, is defined as:

\begin{equation}
\%\text{Gap} = \frac{nRMSE_{\text{proposed}} - nRMSE_{\text{benchmark}}}{nRMSE_{\text{benchmark}}} \times 100.
\label{eq:percentage_gap_nRMSE}
\end{equation}

We utilize this metric to assess the discrepancy between our proposed method and the benchmark, the latter representing the established upper bound for quality. A \%Gap value asymptotically approaching zero indicates an enhanced performance of the proposed approach, nearing this benchmark standard.

\begin{table}[htbp]
    \centering
    \caption{Simulated hour-specific travel time \textit{nRMSE} for case study networks.}
    \label{tab:hourly_eta_nrmse_simulated}
    \begin{adjustbox}{scale=1.0,center}
    \begin{tabular}{lcccccr}
        \toprule
        \textbf{Network} & \textbf{Hour of day} & \textbf{Baseline} & \textbf{Benchmark} & \textbf{Proposed} & \textbf{\%Gap} \\
        \midrule
        \multirow{5}{*}{\textbf{Seattle}} & 13 & 52 & 44 & 45 & 2 \\
        & 14 & 67 & 54 & 54 & 0 \\
        & 15 & 74 & 60 & 64 & 7 \\
        & 16 & 75 & 59 & 62 & 5 \\
        & 17 & 70 & 54 & 54 & 0 \\
        \midrule
        \multirow{5}{*}{\textbf{Orlando}} & 13 & 48 & 46 & 46 & 0 \\
        & 14 & 41 & 39 & 40 & 3 \\
        & 15 & 44 & 41 & 42 & 2 \\
        & 16 & 44 & 40 & 41 & 3 \\
        & 17 & 47 & 41 & 43 & 5 \\
        \midrule
        \multirow{5}{*}{\textbf{Denver}} & 13 & 45 & 44 & 45 & 2 \\
        & 14 & 54 & 53 & 54 & 2 \\
        & 15 & 64 & 55 & 59 & 7 \\
        & 16 & 70 & 53 & 54 & 2 \\
        & 17 & 60 & 50 & 58 & 16 \\
        \midrule
        \multirow{5}{*}{\textbf{Philadelphia}} & 13 & 42 & 41 & 42 & 2 \\
        & 14 & 47 & 42 & 44 & 5 \\
        & 15 & 57 & 52 & 53 & 2 \\
        & 16 & 60 & 55 & 55 & 0 \\
        & 17 & 57 & 51 & 53 & 4 \\
        \midrule
        \multirow{5}{*}{\textbf{Boston}} & 13 & 51 & 50 & 50 & 0 \\
        & 14 & 64 & 53 & 59 & 11 \\
        & 15 & 68 & 56 & 58 & 4 \\
        & 16 & 67 & 53 & 56 & 6 \\
        & 17 & 68 & 48 & 50 & 4 \\
        \bottomrule
    \end{tabular}
    \end{adjustbox}
\end{table}

For off-peak hours and all cities but Seattle, the observed differences in \textit{nRMSE} values between the baseline and benchmark methods are not substantial. This outcome aligns with our earlier observations from the scatter plots and is largely anticipated. As discussed before, this is due to the fact that during off-peak periods, most highway routes experience nearly free-flowing traffic conditions. Hence, underestimated demands still yield representative OD travel times. Seattle, instead, exhibits higher congestion levels even between 1 PM and 3 PM. As a result, we observe a substantial improvement in travel time \textit{nRMSE} for both the benchmark and the proposed methods compared to the baseline.

For peak hours, we observe more significant improvement for both the benchmark and our proposed method in contrast to the baseline. In nearly all cases, the \textit{nRMSE} metric for our proposed method is either marginally worse or the same as that of the benchmark. This is reflected by the \%Gap metric ranging from 0\% to 16\% with a median value of 3\%. This hour-specific travel time \textit{nRMSE} analysis provides quantitative confirmation of the qualitative observations made in Figures \ref{fig:sim_eta_scatter_part1}-\ref{fig:sim_eta_scatter_part3}.

Across all city-hour instances, compared to the baseline, the benchmark achieves an average 13\% reduction in OD travel time \textit{nRMSE}, while our proposed method achieves an average 10\% reduction. Considering only off-peak hours, the average reduction is 8\% for the benchmark and 6\% for our proposed method. Considering only peak-hours, the average reduction reaches 16\% for the benchmark and 13\% for our proposed method.

\begin{figure}[htbp]
    \centering
    \begin{subfigure}[b]{0.48\textwidth}
        \centering
        \includegraphics[width=\linewidth]{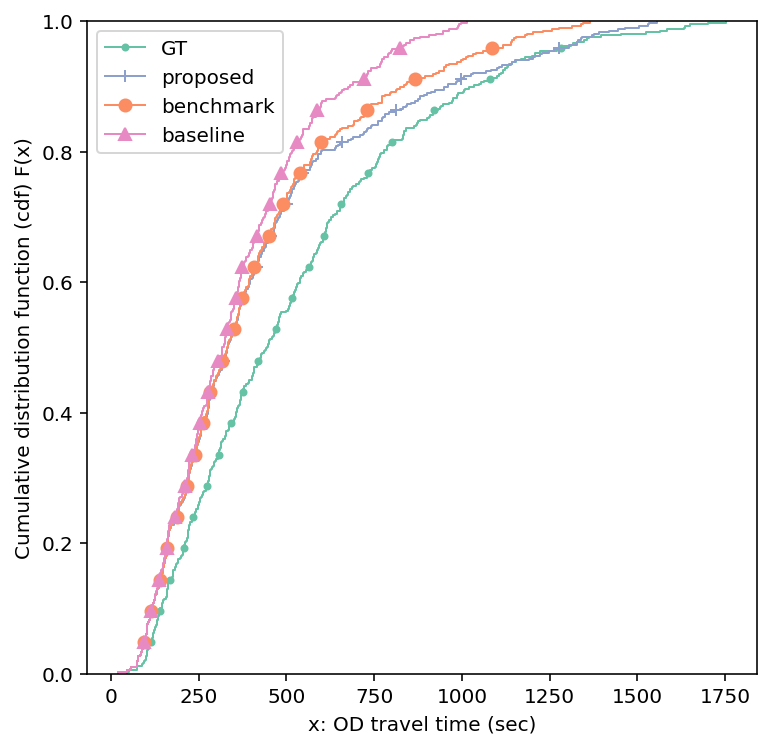}
        \caption{Seattle: Hour 13.}
        \label{subfig:seattle_eta_cdf_1pm}
    \end{subfigure}
    
    \begin{subfigure}[b]{0.48\textwidth}
        \centering
        \includegraphics[width=\linewidth]{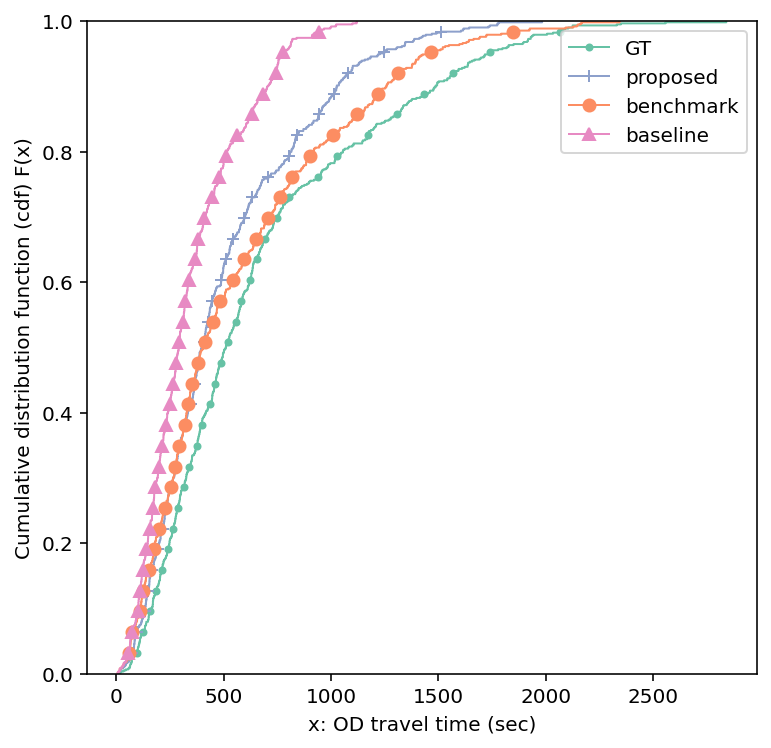}
        \caption{Denver: Hour 16.}
        \label{subfig:denver_eta_cdf_4pm}
    \end{subfigure}
    \hfill 
    \begin{subfigure}[b]{0.48\textwidth}
        \centering
        \includegraphics[width=\linewidth]{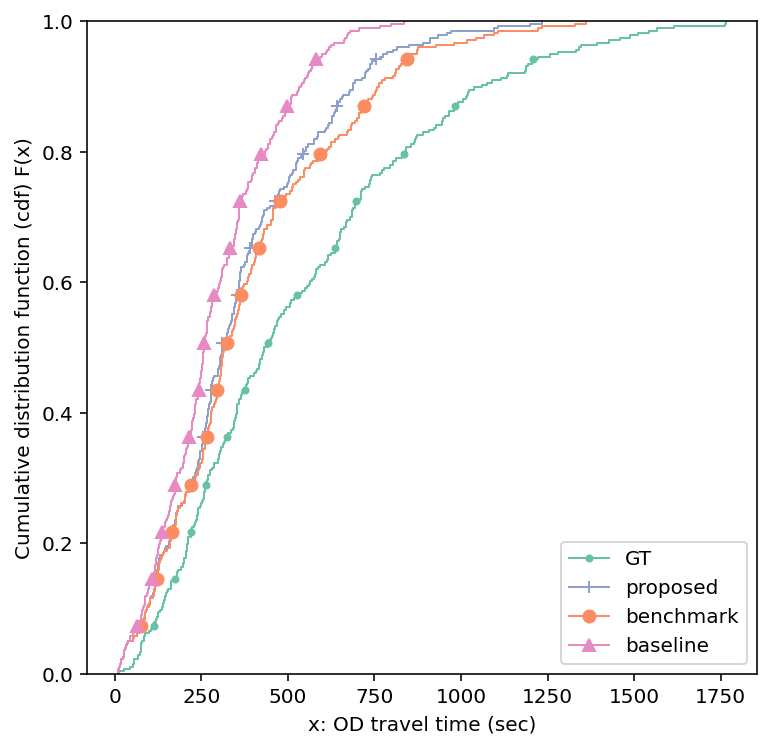}
        \caption{Boston: Hour 16.}
        \label{subfig:boston_eta_cdf_4pm}
    \end{subfigure}

    \caption{Simulated travel time quality analysis for selected case study networks - Cumulative distribution function (cdf).}
    \label{fig:sel_sim_eta_cdf}
\end{figure}

To build on the qualitative insights from scatter plots and the quantitative performance assessment using \textit{nRMSE}, we further analyze a subset of scenarios by visualizing the entire distribution of the OD travel times. Figure \ref{fig:sel_sim_eta_cdf} displays the cumulative distribution functions (cdf(s)) of OD travel time estimates for three selected scenarios: Seattle (hour 13), Denver (hour 16), and Boston (hour 16). Within each subplot, distinct curves represent the GT and three different methods. A given cdf curve can be interpreted as follows: the degree to which a given curve approximates the GT curve (the blue line) reflects the goodness-of-fit of its corresponding simulated OD travel times to the empirical distribution. More specifically, the \textit{y}-value on the cdf curve at a given \textit{x}-value denotes the fraction of ODs with simulated travel times less than or equal to \textit{x}.

The cdf analysis further reinforces our previous findings. The consistent rightmost position of the GT curve and leftmost position of the baseline cdf systematically confirm the baseline's underestimation of travel times, underscoring the critical need for demand upscaling. The benchmark's cdf closely approximates the GT, highlighting its high-resolution modeling accuracy, though minor discrepancies suggest other influencing factors that are not accounted for. These may include, but are not limited to, GT data quality, simulation calibration accuracy, and assumptions regarding uniform scaling. Crucially, our proposed method consistently outperforms the baseline, with its cdf curves robustly positioned between the GT and the baseline. Moreover, it often aligns closely with the benchmark, demonstrating near-benchmark performance despite its parsimonious analytical approach.

Our proposed analytical demand upscaling method offers significant value by consistently achieving improved travel time fitting comparable to a high-resolution benchmark, but without its substantial computational cost. This makes it scalable across cities worldwide. It also makes it appropriate for large-scale network applications. The method's consistent performance across diverse highway networks further demonstrates its scalability and robustness.

In conclusion, the ODs derived by the proposed method yield realistic OD travel times, consistently approaching the performance of a high-fidelity benchmark. A key advantage is its exceptional computational efficiency, leveraging a tractable analytical network model, which enhances scalability and makes it a practical tool for resource-constrained applications.

\section{Discussion and Conclusions}
\label{sec:cl}
This paper introduces and validates a worldwide scalable, and fast to compute, method to estimate travel demand from incomplete aggregated traffic statistics. The proposed technique systematically infers a complete demand matrix by upscaling a representative subsample demand, leveraging information derived from aggregated observed travel times. The determination of the optimal upscaling factor is achieved through an analytical formulation derived from an inexpensive macroscopic network model, which boasts low calibration overhead and rapid evaluation capability. The problem is formulated as a one-dimensional optimization problem, characterized by a quadratic objective function and subject to nonlinear and bound constraints. It is implemented as bound-constrained problems with a nonlinear (and non-quadratic) objective function that can be solved efficiently. 

Validation is performed on the large-scale highway networks of Los Angeles and San Diego. To carry out a robust validation, we use a new type of data for validation: sensor segment count data. We apply the approach hourly for 5 AM - 8 PM of a weekday and for the two networks. The approach achieves improvements in count \textit{nRMSE}: 64\% for Los Angeles and 74\% for San Diego, significantly outperforming a baseline approach that solely utilizes unscaled demand statistics. Further case studies on five large-scale highway networks consistently illustrate the efficacy and scalability of our proposed method. Analyzing data from 1 PM to 6 PM in each city, our approach yields an overall 10\% improvement in travel time \textit{nRMSE} compared to the baseline. This improvement is even more pronounced, reaching 13\% for a subset of more congested hours (i.e., afternoon peak). Moreover, our proposed method consistently performs comparably to a high-fidelity, simulation-driven benchmark approach, with a median quality gap of just 3\% across all scenarios, underscoring its practical value in replicating congested traffic patterns.

As discussed earlier in the paper, there are numerous avenues for future research. Of particular interest, is the extension of the proposed approach to replicate travel patterns in urban, non-highway, areas. This would rely on the use of fundamental diagrams for urban segments. Recent formulations may be appropriate here \citep{Li22, Zhang22fd, Vishnoi23}. The approach currently solves OD estimation problem separately for each time interval of interest (in our case, every hour). Of interest would be to jointly solve the problem across time intervals, while also enforcing a regularity constraint that describes how the scaling factor smoothly varies across time. Similarly, the formulation of a model that describes the spatial variation of the scaling factor, while ensuring a smooth variation, would give the model more degrees of freedom, allowing it better capture the spatial variations of congestion in the networks.

\bibliographystyle{elsarticle-harv} 
\bibliography{biblio_spacetime}

\end{document}